\documentclass[12pt]{mn2e}
\usepackage{myaasmacros,epsfig, amssymb}        

\title[EROs]{The Las Campanas Infra-red Survey. V.\ Keck Spectroscopy of
a large sample of  Extremely Red Objects}
\author[M.Doherty et al.]
  {M.~Doherty,$^1$\thanks{email:md@ast.cam.ac.uk}
  A.J.~Bunker,$^2$ R.S.~Ellis$^3$ and P.J.~McCarthy,$^4$ \\
  $^1$Institute of Astronomy, University of Cambridge, Madingley Road, Cambridge, CB3\,0HA, U.K.\\
  $^2$School of Physics, University of Exeter, Stocker Road, Exeter, EX4\,4QL, U.K.\\
  $^3$California Institute of Technology, Astronomy, Mail Stop 105-24, Pasadena, CA~91125, U.S.A.\\
  $^4$Observatories of the Carnegie Institute of Washington, Santa Barbara Street, Pasadena, CA~91101, U.S.A.\\
}

\date{Released 2005 Xxxxx XX}

\pagerange{\pageref{firstpage}--\pageref{lastpage}} \pubyear{2005}

\def\LaTeX{L\kern-.36em\raise.3ex\hbox{a}\kern-.15em
    T\kern-.1667em\lower.7ex\hbox{E}\kern-.125emX}

\def\Ha{\ifmmode \mathrm{H}{\alpha}\else H$\alpha$\fi}
\def\micron{\ifmmode {\mu}\mathrm{m}\else $\mu$m\fi}
\def\msol{\ifmmode \mathrm{M}{_{\odot}}\else M$_{\odot}$\fi}
\newcommand {\apgt} {\ {\raise-.5ex\hbox{$\buildrel>\over\sim$}}\ }
\newcommand {\aplt} {\ {\raise-.5ex\hbox{$\buildrel<\over\sim$}}\ }

\begin{document}

\label{firstpage}

\maketitle

\begin{abstract} 
  
We present deep Keck spectroscopy, using the DEIMOS and LRIS spectrographs, of a
large and representative sample of 67 ``Extremely Red Objects'' (EROs) 
to $H=20.5$ in three fields (SSA22, Chandra Deep Field South and the NTT Deep Field)
drawn from the Las Campanas Infrared Survey. Using the colour cut $(I-H)>3.0$ 
(Vega magnitudes) adopted in earlier papers in this series, we verify the efficiency 
of this selection for locating and studying distant old sources. Spectroscopic
redshifts are determined for 44 sources, of which only two are contaminating
low mass stars. When allowance is made for incompleteness, the spectroscopic 
redshift distribution closely matches that predicted earlier on the basis of photometric 
data. Our spectra are of sufficient quality that we can address the important question 
of the nature and homogeneity of the $z>$0.8 ERO population. A dominant old stellar 
population is inferred for 75\% of our spectroscopic sample; a higher fraction that 
than seen in smaller, less-complete  samples with broader photometric selection 
criteria (e.g. $R-K$). However, only 28\% have spectra with no evidence of recent 
star formation activity, such as would be expected for a strictly passively-evolving
population. More than $\sim$30\% of our absorption line spectra are of the `E$+$A' type 
with prominent Balmer absorption consistent, on average, with mass growth of 5-15\% in 
the past Gyr. We use our spectroscopic redshifts to improve earlier estimates of the 
spatial clustering of this population as well as to understand the significant 
field-to-field variation. Our spectroscopy enables us to pinpoint a filamentary 
structure at $z=1.22$ in the Chandra Deep Field South. Overall, our study
suggests that the bulk of the ERO population is an established population
of clustered massive galaxies undergoing intermittent activity consistent with
continued growth over the redshift interval 0.8$<z<$1.6

\end{abstract} 


\section{Introduction}
\label{sec:intro}

One of the major challenges facing the hierarchical models of galaxy
formation (e.g. Somerville et al. 2004\nocite{smm+04}) is posed by the
large number of massive galaxies with apparently well-established stellar
populations observed at high redshift (Glazebrook et al.\ 2004, Cimatti et al.\
2004\nocite{gam+04}\nocite{cdr+04}). Galaxies with evolved stellar populations at 
moderate redshifts ($z\sim1-2$) can be characterized by very red optical-near-infrared
colours. The first examples were identified shortly 
after the introduction of near-infrared imaging arrays (Elston, Rieke \& Rieke, 
1988; McCarthy, Persson, \& West 1992\nocite{mpw92}; Cowie et al.\ 1990\nocite{cglm90}).  Spectroscopy of two 
of the Elston et al. (1988)\nocite{err88} red candidates revealed them to be luminous evolved
galaxies at $z\sim0.8$ (Elston et al.\ 1989)\nocite{err89}.  More distant red galaxies
with evolved populations were later identified from deep radio surveys (Dunlop
et al.\ 1996\nocite{dps+96}). 

The optical-to-near-infrared colour selection appears to yield a mixture of
evolved systems and actively star-forming galaxies with substantial
internal reddening (see McCarthy 2004 for a review). The prototype of the
latter class (ERO J164502+4626.4) was identified by its extremely red $R-K$ colour by Hu \&
Ridgway (1994)\nocite{hr94}.  Subsequent spectroscopy and sub-mm
observations revealed this system to be a star-forming galaxy at
$z=1.44$ (Graham \& Dey 1996\nocite{gd96}; Cimatti et al.\ 1998, Dey et
al.\ 1999\nocite{cart98}\nocite{dgi+99}).

Earlier papers in this series (McCarthy et al 2001\nocite{mcc+01}, Firth et al.\ 2002; 
Chen et al.\ 2002) presented source counts, angular clustering and photometric 
redshifts for objects selected via optical-near infrared colours using the
Las Campanas Infra-red Survey (LCIRS) -- one of the first to use a panoramic
near-infrared camera. This work has been complemented by independent surveys 
of $K$-band selected samples (e.g. Cimatti et al.\, 2002  -- K20; Yan et
al. 2003\nocite{yts03} ; 
Abraham et al.\, 2004\nocite{agm+04} -- GDDS). Although the abundance of EROs is
less than that implied if all of the present-day spheroidal galaxies
followed a passive evolutionary track from high redshift, it is much higher 
than that predicted by current semi-analytical models (Firth et al. 2002\nocite{fsm+02}; see also Cimatti et al. 2002\nocite{cdm+02}).  

Strong clustering has been observed for the ERO population, with amplitudes up to 10 
times that seen in the equivalent flux-limited sample (see e.g. Daddi et al. 2000, Firth et al. 2002). On the basis of
photometric redshifts, Firth et al.\ (2002) and McCarthy et
al.\ (2001) derived co-moving correlation lengths of $r_o\approx
6-10h^{-1}$\,Mpc (where h=H$_0$/100~km~s$^{-1}$~Mpc$^{-1}$), comparable to that for early-type galaxies at low redshift. 
Daddi et al. (2002)\nocite{dcb+02} infer a similar, although somewhat larger, clustering scale 
on the basis of Passive Luminosity Evolution (PLE) models of the redshift 
distribution of the K20 sample. 

\begin{figure*}
\begin{tabular}{cc}
\psfig{figure=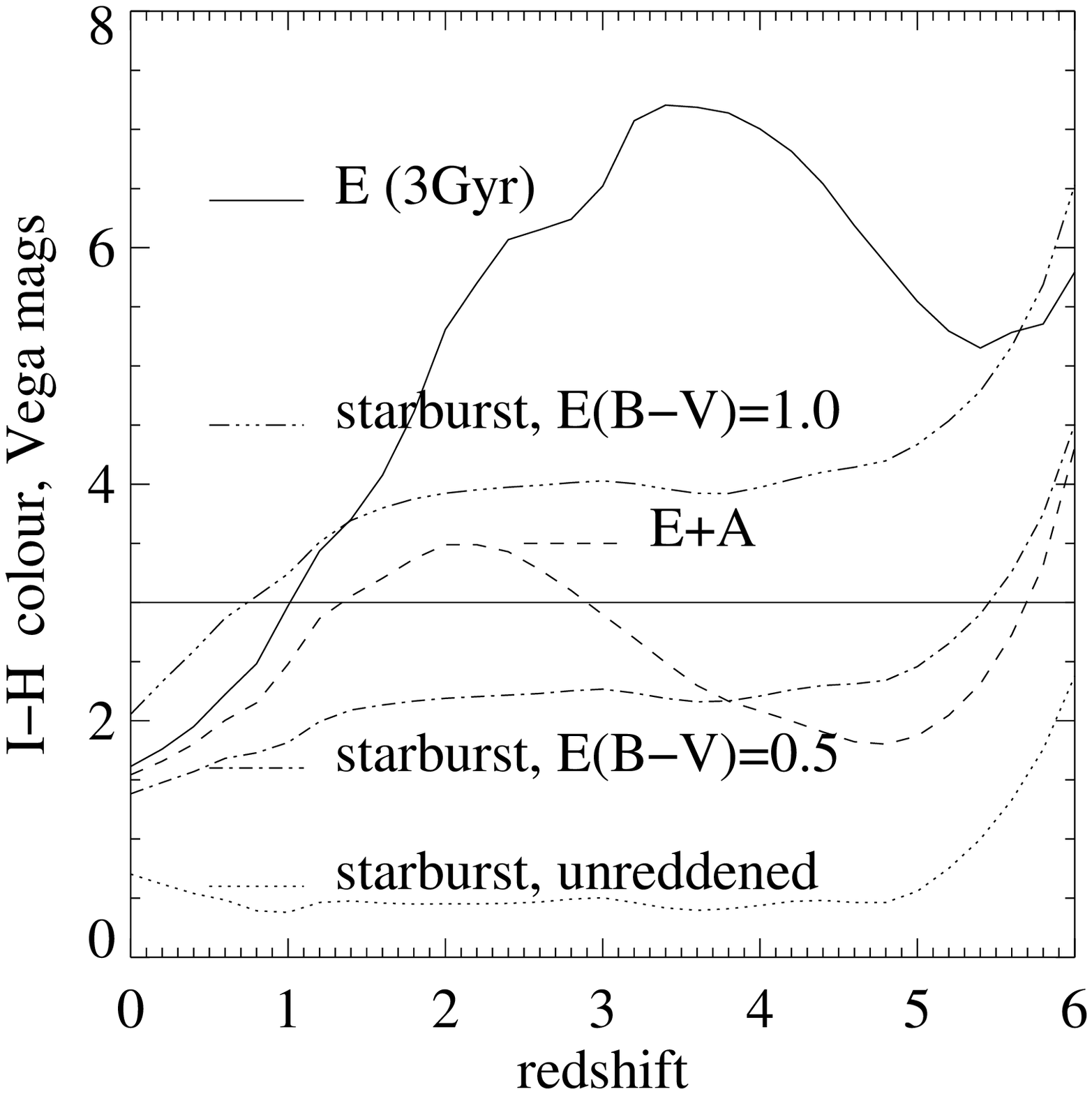,width=80mm} &
\psfig{figure=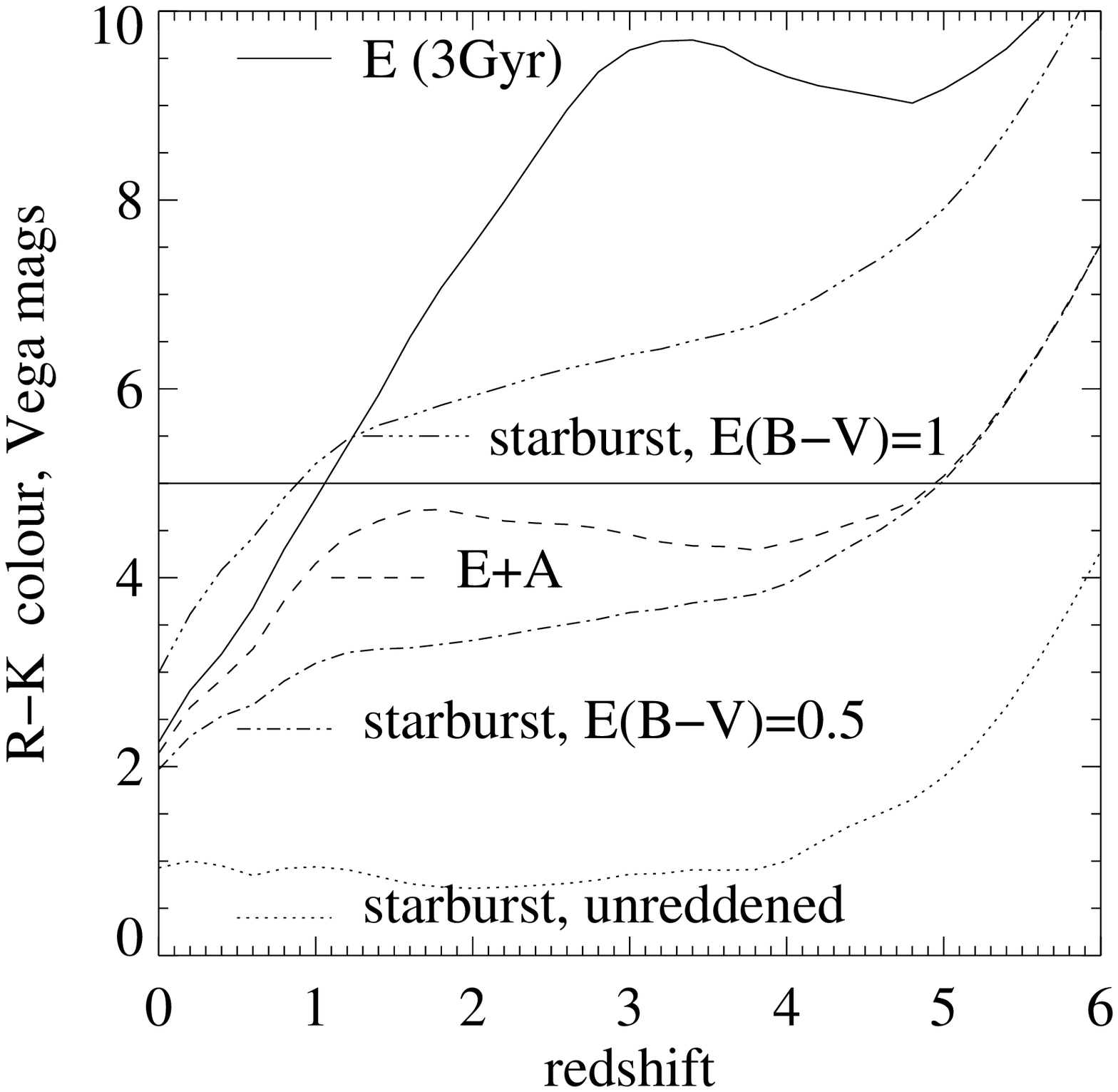,width=80mm} 
\\
{\bf (a)} & {\bf (b)} \\
\end{tabular}
\caption{(a) Evolution in the predicted ($I-H$) colours (Vega system) for various 
stellar populations, produced with Bruzual \& Charlot's (2000) synthesis code
assuming a Scalo (1986) initial mass function. The elliptical (E) spectral 
energy distribution (solid line) has an age of 3~Gyr. The E+A spectrum is
equivalent but viewed 100~Myr after a burst of star formation involving 5\% of 
the stellar mass. The starburst galaxy assumes an exponential star formation
history with an e-folding time of 1~Gyr, viewed 10~Myr after the onset of star 
formation. Reddening of $E(B-V)=0.5$ and $1.0$ have been applied using
Calzetti's (1997) prescription. The $I-H>3$ colour criterion (shown)
is designed to select passive objects with redshifts $z>1$ and highly
reddened starbursts with $z>0.7$. For comparison, the equivalent tracks for
an $R-K>5$ colour cut are shown in the adjacent panel (b). The essential
difference evident here is that E$+$A type galaxies would not be expected to occur in an
$R-K>5$ selected sample.}

\label{fig:tracks}
\end{figure*}

Spectroscopic data has played a key role in furthering our understanding
of the ERO population. Dunlop et al.\ (1996) and \nocite{sds+97}Spinrad et al.\ (1997) 
used low resolution spectroscopy to infer an age in excess of 3.5~Gyr for 
the radio-selected red galaxy LBDS 53W091 at $z \sim 1.5$. More recent intermediate
dispersion spectra for co-added samples of fainter colour-selected objects 
at $1.3 < z < 2$ also point towards early formation redshifts for a 
significant fraction of the red population (McCarthy et al.\ 2004\nocite{mlc+04}; 
Cimatti et al.\ 2004).

The goal of this paper is to build on the earlier LCIRS photometric studies
of EROs, using deep Keck spectroscopy. We investigate the redshift distribution and 
spectroscopic properties of
a representative subset of the ERO population. 
In order to verify and further exploit the earlier work, our spectroscopic
targets were selected from three of the LCIRS fields according to the 
same $I-H$ colour cut (McCarthy et al.\ 2001). 

With the resulting spectroscopic sample we aim to: 
\renewcommand{\theenumi}{\roman{enumi})}
\begin{enumerate}
\item verify that the $I-H$ colour selection is optimal for selecting passively 
evolving galaxies at $z>1$, 
\item determine the redshift distribution of the population and compare
it with that determined photometrically,
\item examine the spectroscopic nature of EROs with respect to the
(perhaps idealized) view that such a colour criterion locates a passively-evolving
population of sources destined to emerge as present-day spheroidal galaxies.
\end{enumerate}

It should be borne in mind that spectroscopic (or morphological) studies
can present an incomplete picture of the evolutionary history of a given 
population. The age of a stellar population need not necessarily coincide
with the age of the assembled mass. In particular, a spectral classification
indicative of an established or aged stellar population need not
necessarily correspond to a morphological elliptical. However, spectral
diagnostics of recent activity can be invaluable indicators of 
continuing growth. Our purpose in this paper is largely to explore the
spectroscopic {\it homogeneity} of the ERO population as well as the
extent to which the idealized picture of a passively-evolving component
can be made to fit the data. We will examine and compare the morphological 
information on a similar population in the next paper in this series 
(Doherty et al. {\it in prep.}) 

A plan of the paper follows: in Section 2 we present the sample
selection and in Section 3 we detail the Keck spectroscopic observations and 
their reduction. Section 4 presents the inferred redshift distribution taking into 
account incompleteness effects. We analyze our results in terms of our primary 
objectives (above) in Section 5. Our conclusions are presented in Section 6.  

Unless otherwise stated, throughout the paper we use the standard
``concordance'' cosmology of $\Omega_M=0.3$, $\Omega_{\Lambda}=0.7$,
and $H_0=70\,{\rm km\,s^{-1}\,Mpc^{-1}}$. All magnitudes are 
on the Vega system.

\section{Sample selection}
\label{sample-sel}
Our selection of Extremely Red Objects is drawn from three (out of a
possible five) LCIRS fields at high Galactic latitude accessible at the 
Keck observatory: the Chandra Deep Field South (CDFS; Giacconi et
al.~2001\nocite{gzw+02} ), the NTT Deep Field (Arnouts
et~al. 1999\nocite{adc+99} ) and SSA22 (part of the 
CFRS redshift survey, Lilly et al.~1995\nocite{llch+95} ). 

As discussed, we employ the same colour criterion ($I-H>3$) for EROs as that 
adopted in the earlier photometric study by McCarthy et
al. (2001). Figure~\ref{fig:tracks} demonstrates that this colour cut is
designed to select passive objects with redshifts $z>1$ and starbursts with
substantial dust reddening. To
facilitate a reasonable success rate with optical spectroscopy, we adopted
a magnitude limit of $H=20.5$. At this limit, the LCIR survey is close to 100\% complete (see discussion in Chen et al. 2002 -- \S4.3 and Figure 5) and galaxies with $I-H>3$ comprise roughly 
$\sim10\%$ of the total population, with a surface density of $\approx
0.8$\,arcmin$^{-2}$. 

We briefly recap the photometric data which forms the basis of our sample.
The infrared imaging was obtained with the CIRSI camera (Beckett et al. 1998)\nocite{bmm+98} on
the Du Pont 2.5~m 
telescope at Las Campanas Observatory
between 1998 and 2000. CIRSI has 4 HgCdTe $1024^2$ infrared (IR) arrays, spaced
by 0.9 array widths, with a pixel scale of $0.2$\arcsec.  By
observing a $2\times 2$ mosaic, a contiguous 13$\times$13arcmin
(170\,arcmin$^2$) field was surveyed.  In the case of SSA22, only
imaging data from half the area was reduced in time to select EROs for
LRIS mask manufacture for our October 2001 observing run. Hence the
total survey area from which we built catalogues of EROs is
425\,arcmin$^2$. The photometric data reduction is detailed in Firth et al.\ 
(2002)\nocite{fsm+02} and Chen et al.\ (2002)\nocite{cmm+02}. In the
reduced $H$-band images, the typical $5\,\sigma$ detection
is $H=21.5$ in a 3\arcsec-diameter aperture. The optical
imaging came from the CFHT 12k Mosaic ($V$-, $R$- \& $I$-band for
SSA22), the Wide Field Camera on the 2.5\,m Isaac Newton Telescope
(for NTT), and the Mosaic-II camera on the CTIO 4\,m ($V$-, $R$- $I$-
\&$z$-band for CDF-S). The optical and infrared images were registered and 
distortion-corrected to an astrometric frame determined using the Digitized 
Sky Survey. Colour selection was performed using version 2.2.1 of the 
SExtractor package (Bertin \& Arnouts 1996) in two-image mode with a fixed 
aperture of 3\arcsec diameter.  

Spectroscopy was performed at Keck with both the Low Resolution Imaging Spectrograph 
(LRIS; Oke et al. 1995\nocite{occ+95}) and the Deep Imaging Multi-Object Spectrograph 
(DEIMOS, Faber et al.\ 2003; Phillips et al.\ 2002)\nocite{pfk+02}\nocite{fpk+03}. 
The $5'\times 7'$ LRIS slitmask field is much smaller than a CIRSI tile, whereas
the DEIMOS field covers $16.5'\times 5'$ (i.e., about half the area surveyed in a 
CIRSI tile). In selecting sources for multi-object spectroscopy, the astrometric
centre and position angle of the masks were optimised within the LCIRS fields 
to maximize the number of candidate EROs. We discuss later the possible
biases that this may bring to our analyses. The minimum slit length was 
set to 6\arcsec, and the selection was determined entirely by geometric
constraints. In the event of a slit clash higher priority was given to the 
brighter source (in $I$). 

\begin{figure}
\begin{tabular}{c}
CDFS \\
\psfig{figure=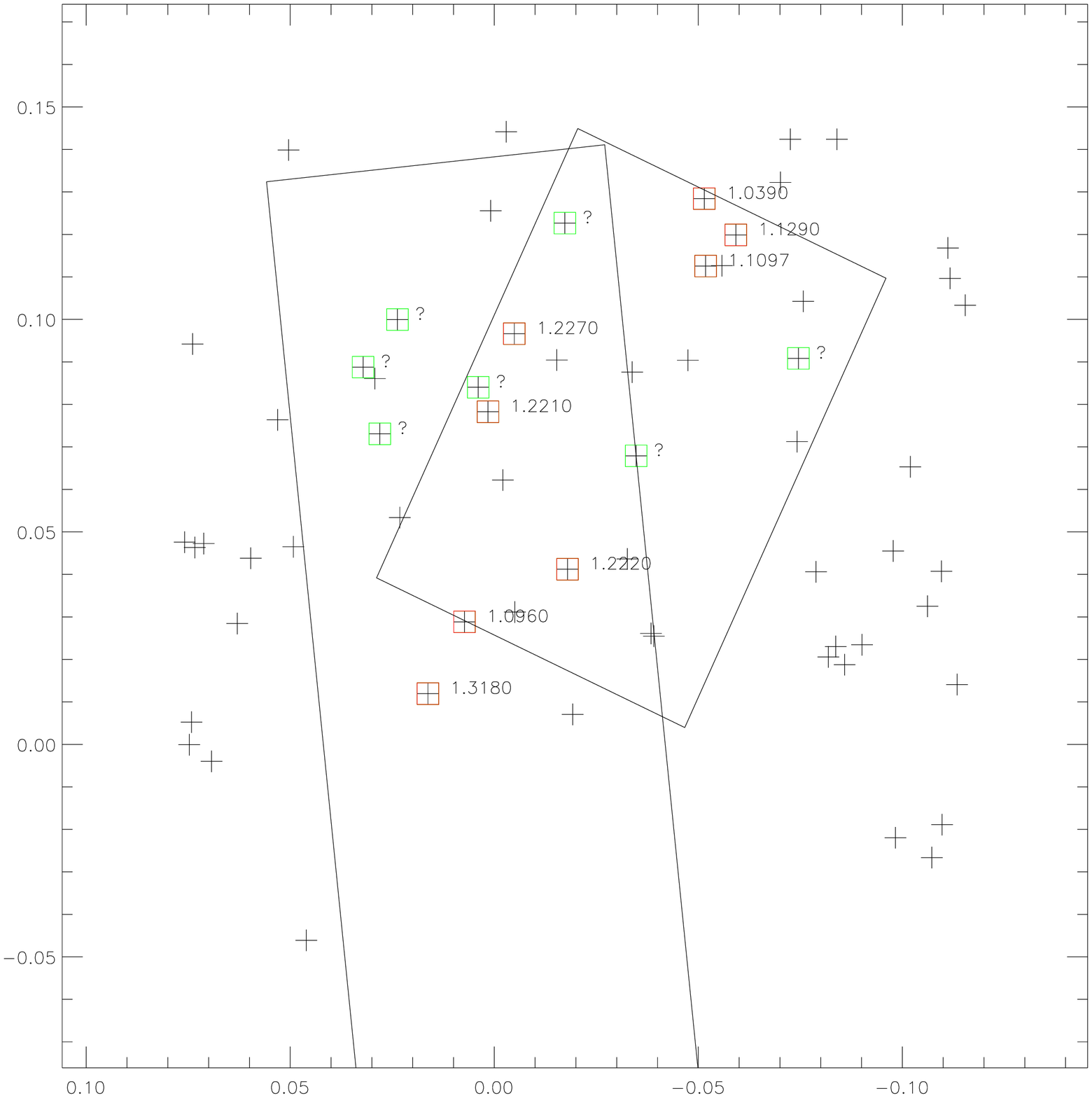,width=75mm} \\
SSA22 \\
\psfig{figure=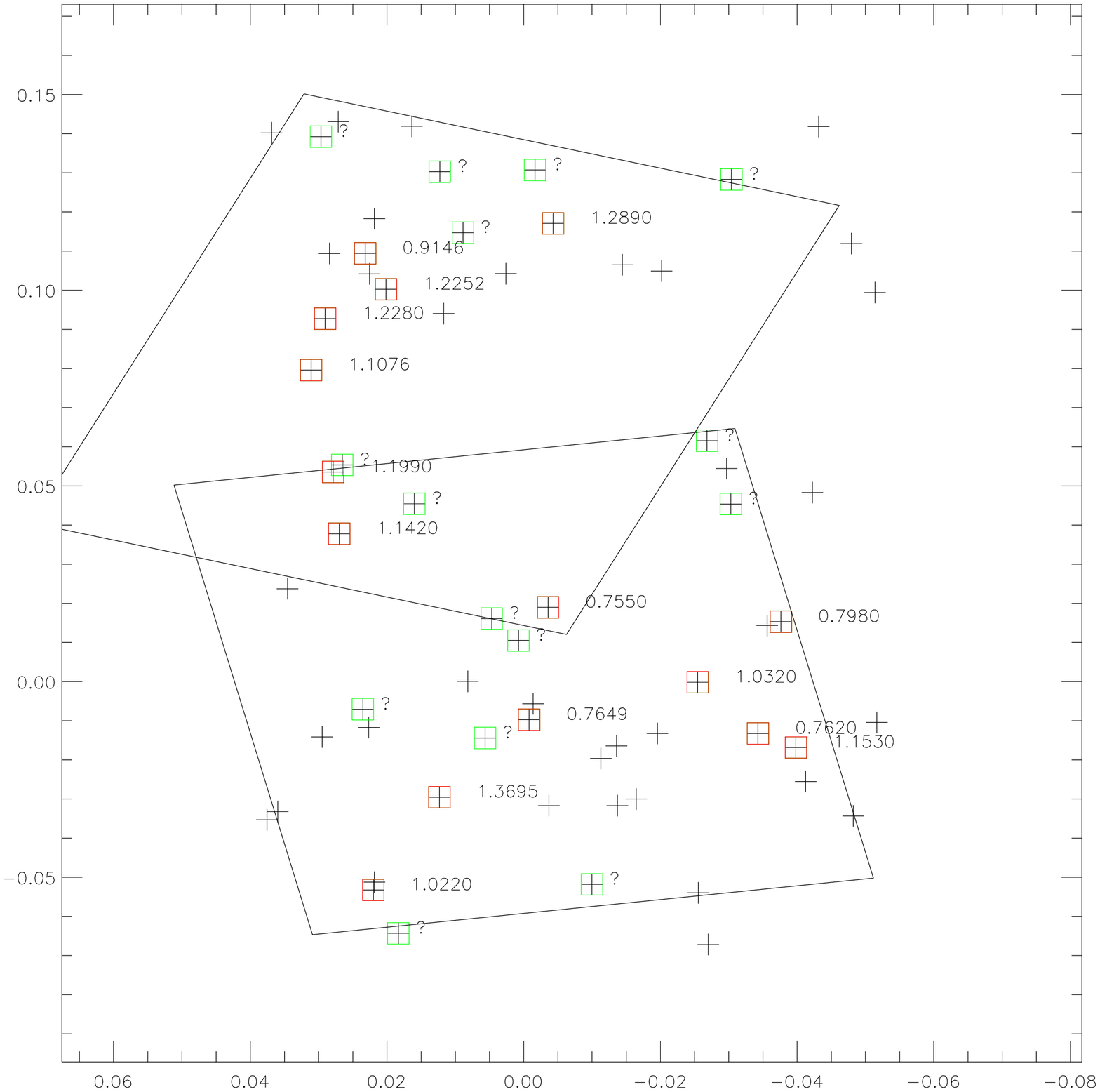,width=75mm} \\
NTT \\
\psfig{figure=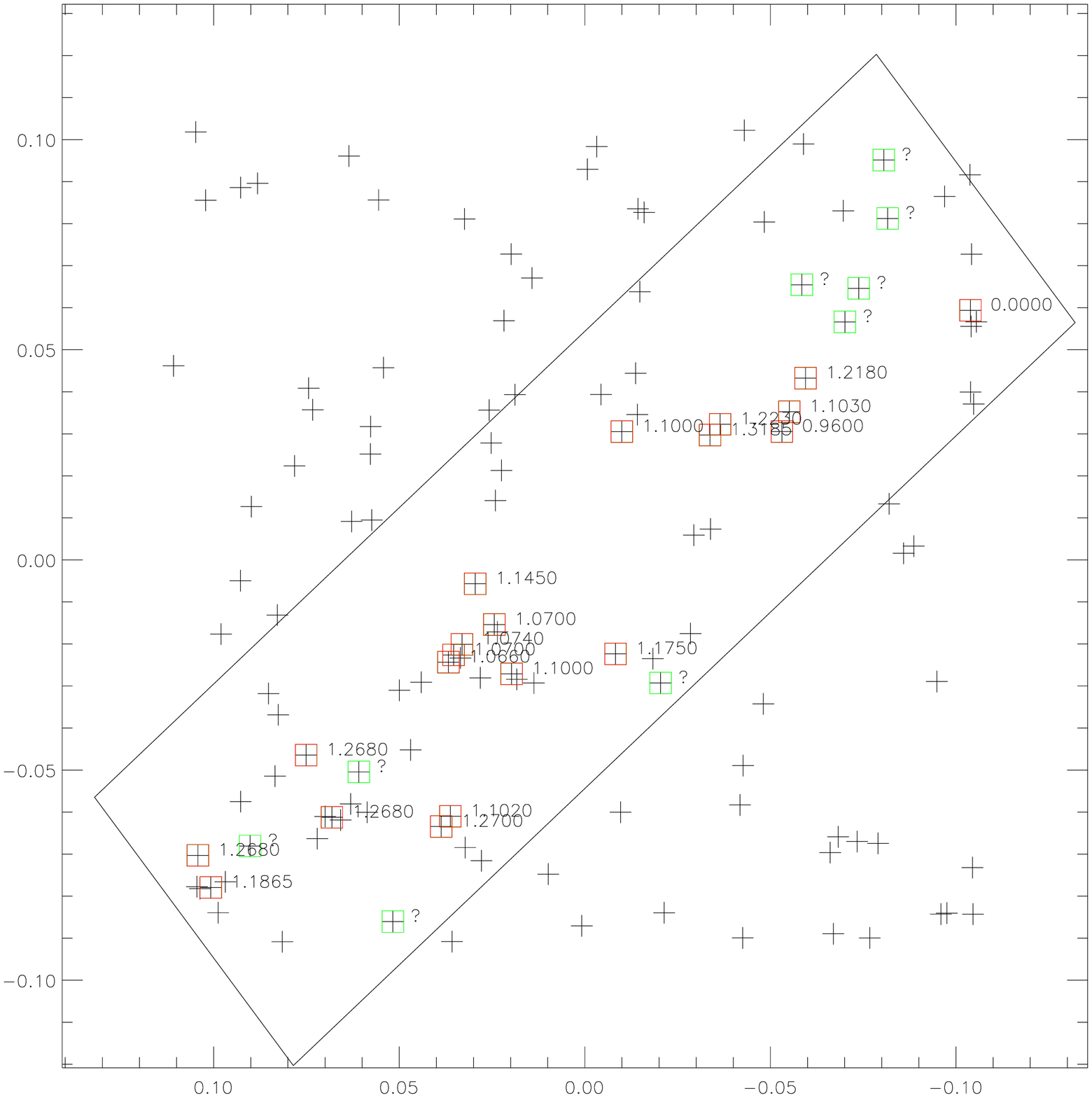,width=75mm} \\
\end{tabular}
\caption{Distribution of EROs for the three LCIRS fields. Candidates
  targetted spectroscopically are marked with boxes. Where redshifts were 
  identified the boxes are marked accordingly (otherwise indicated '?'). 
  The areas delineated refer to the spectroscopic footprints of the slit-mask.}
\label{fig:sky-distrib}
\end{figure}

The fraction of EROs within the field of view of each spectrograph
that could be incorporated onto a mask was roughly 50\%
(29/73 in NTT, 16/31 in CDFS, 30/55 in SSA22). The angular distribution of
EROs is shown in Figure~\ref{fig:sky-distrib}. Part of the DEIMOS mask
for CDFS was taken up with targets for another program involving
spectroscopy of $i'-$band drop-out redshift $\sim6$ galaxies (see
Bunker et al.\ 2003\nocite{bse+03}; Stanway et al.\ 
2004\nocite{sbm+04}), which necessitated part of the mask falling
outside of the LCIRS field. The spectroscopically surveyed area from
each LCIRS image is thus: $\approx 80\,{\rm arcmin}^{2}$ for NTT;
$\approx 60\,{\rm arcmin}^{2}$ for CDFS; and $\approx 60\,{\rm
arcmin}^{2}$ for SSA22, a total of about $200\,{\rm arcmin}^{2}$.

\section{Spectroscopy}
\label{sec:spectroscopy}
\subsection{LRIS multi-object spectroscopy}
\label{subsec:lrisobs}

\begin{table*}
\begin{tabular}{cccccccc}
\hline
Field & Field Centre (J2000) &Instr. &date/time& total exptime &PA (deg)&slit
width (arcsec) &No.\ targets \\ 
\hline
SSA22\#1 & 22:17:41.9 00:13:53 & LRIS & 22--24 Oct 2001 & 16ks & 10&1.5 &16\\
SSA22\#2 & 22:17:45.0 00:18:45 & LRIS & 22--24 Oct 2001 & 26ks &-20 & 1.0&12\\
CDFS    & 03:32:26.1 -27:43:24 & LRIS & 22--24 Oct 2001 & 18ks & -25
& 1.5&11\footnote{6 of these objects were also observed with DEIMOS}\\
CDFS    & 03:32:35.2 -27:47:52 & DEIMOS & 8-9 Jan 2003 & 20 ks & 6& 1.0
&15\footnote{including 6 objects on the LRIS mask and 6 objects selected according to $R-K>5$} \\
NTT     & 12:04:10.2 -07:26:06 &  DEIMOS & 8-9 Jan 2003 & 19.2 ks & -50&1.0 &23 \\

\hline

\end{tabular}
\caption{Summary of Spectroscopic Observations}
\label{tab:obs}
\end{table*}

We obtained long-slit spectroscopy of 41 EROs on the nights of 2001
October 22--24 UT using LRIS (Oke et al. 1995\nocite{occ+95}) at the f/15 
Cassegrain focus of the 10-m Keck{\scriptsize~I} telescope.  Spectra 
were obtained for 11 EROs in CDFS (including 6 which overlapped with the 
later DEIMOS sample, Section~\ref{sec:deimosobs}), and for 30
EROs in SSA22 (using 2 different masks). Full details are given in
Table~\ref{tab:obs}.

The LRIS red-arm detector is a Tek 2048$^2$ CCD with 24\,\micron\ 
pixels.  The angular scale is 0\farcs 212\,pixel$^{-1}$, and the CCD
was read out in two-amplifier mode. Observations were obtained
using the 600~line~mm$^{-1}$ grating in first order blazed at
7500\,\AA, producing a dispersion of 1.24\,\AA\,pixel$^{-1}$. The
reference arc lamps and sky-lines have a full width at half maximum
(FWHM)~$\approx$~5--8\,\AA\ (the spectral focus being best at the
central wavelength). For objects that fill a 1\arcsec-wide slit,
the velocity width of a spectrally unresolved line is FWHM~$\approx
220-300$\,km\,s$^{-1}$. A wider 1.5\arcsec\ slit was used for CDFS and
SSA22\#1, producing slightly lower spectral resolution
($\sim400$km~s$^{-1}$ FWHM). The grating was tilted to place a central
wavelength of $\lambda=8200$\,\AA\ on the detector (for slits in the
centre of the mask), sampling the wavelength range $7000-9000$\,\AA\ 
in all targets, with up to $\approx 500$\,\AA\ either end depending on
the position of the slit on the mask.  We simultaneously obtained
exposures with the blue arm of LRIS (read in one-amplifier mode, with
the 300\,lines\,mm$^{-1}$ grating blazed at 5000\,\AA ), using a
dichroic beam-splitter at 6800\,\AA .  However, due to the red nature
of our objects, there was little or no flux at $\lambda \lesssim 6000$~\AA .

The total integration times are given in Table~\ref{tab:obs}. These were
broken into individual exposures of duration 2000\,s to enable
more effective cosmic ray rejection.  The telescope was dithered by $\pm
3$\arcsec along the slit between integrations to facilitate the removal of
fringing in the red and the elimination of bad pixels.  The observations
spanned an airmass range of $1.00-1.5$. The seeing had a FWHM~=~0\farcs
6--1\farcs 0 over the course of the night.  Spectrophotometric standard
stars HZ~4, G191B2B \& Feige~110 (Massey et al. 1988; Massey \& Gronwall
1990\nocite{msba88}\nocite{mg90}) were observed at similar airmass to determine the sensitivity
function for flux calibration.

Each frame was bias subtracted according to values determined from the overscan
region and converted to electrons by the gain appropriate to each amplifier. A 
high signal-to-noise averaged dark current frame was then subtracted.  
Normalized flat fields were derived from exposures taken with 
a halogen lamp immediately after the science exposures. For each dither
position the average of the other two positions was subtracted, to remove
the fringing pattern (which is approximately fixed in space and does not
vary significantly with time), while preserving the object signal. Residual
sky subtraction was 
performed by fitting a low-order polynomial to each detector column (parallel to 
the slit).  Individual nod positions were registered and combined using a 
cosmic ray rejection algorithm, and data on the same mask from different
nights were added using inverse-variance weighting.

One-dimensional spectra were extracted using the {\tt IRAF} {\em apall}
package: extraction widths of 6\,pixels (1.3\arcsec) were used, and the
curvature of individual object spectra was ``traced'' across the array.
Wavelength calibration was obtained from Ne$+$Ar reference
arc lamps, and a fourth-order polynomial fit to the centroids of 40 arc
lines created a wavelength solution with {\em rms} residuals of 0.3\,\AA.

\subsection{DEIMOS multi-object spectroscopy}
\label{sec:deimosobs}
We also obtained slit-mask spectra of 38 EROs using the DEIMOS spectrograph
(Faber et al.\ 2003; Phillips et al.\ 2002)\nocite{pfk+02}\nocite{fpk+03}) at 
the Nasmyth focus of the 10-m Keck{\scriptsize~II} telescope. We used one 
slitmask each on the CDFS field and the NTT field, targeting EROs within 
the $16.5\times5$\,arcmin DEIMOS field. Details are given in Table~\ref{tab:obs}. 
Each slit was 1\arcsec wide, with a minimum slit length of 6\arcsec. 
DEIMOS has 8 MIT/LL $2k\times 4k$ CCDs with $15\,\mu$m pixels and an angular 
scale of 0.1185\arcsec\,pix$^{-1}$.

Observations were obtained using the Gold 1200\,line\,mm$^{-1}$ grating
in first order tilted to a central wavelength of 8400 \AA\ with a dispersion of
$0.32$\,\AA\,pixel$^{-1}$. Each spectrum spanned $\approx 2600$\,\AA\
and for all targets covered $\lambda\lambda_{\rm obs}\,7000-9000$\,\AA. 
A small 8\,\AA\ region in the centre of the wavelength range falls in the
gap between two CCDs. We used the OG550 order-blocking filter to remove all
light at wavelengths short-ward of 5500\,\AA , to avoid contamination 
by second-order light.

The spectral resolution is $\Delta\lambda_{\rm FWHM}^{obs}\approx
1.4$\,\AA\ ($\Delta v_{\rm FWHM}\approx 55$ km s$^{-1}$), as measured
from the sky lines. The seeing was typically $0\farcs6-0\farcs9$ FWHM, i.e. smaller
than the slit width, implying that the resolution quoted above is
a slight overestimate.

The seeing was typically in the range
$0\farcs6-0\farcs9$ FWHM, smaller than the slit width of $1\farcs0$.
As the seeing disk was smaller than the slit width the
true resolution is somewhat better for a source which does not fill
the slit.

Observations were divided into individual exposures of duration
2400\,s. The telescope was dithered 1.5\arcsec along the slit
between integrations.  Flux calibration and telluric correction were
carried out using observations of the spectrophotometric standard
HZ\,44 (Massey et al.\ 1988; Massey \& Gronwall 1990). The flux
calibration was checked using the spectra of the five alignment stars
of known broad-band photometry ($I\approx 17-19$\,mag), which were
used to position the masks through $2''\times2''$ alignment boxes.
Wavelength calibration was obtained from Ne$+$Ar$+$Hg$+$Kr reference
arc lamps.

Most of the spectra were reduced using v1.1.4 of the DEEP2 data
pipeline\footnote{{\tt
    http://astron.berkeley.edu/$\sim$cooper/deep/spec2d/}}.  The
pipeline rectifies the slitlets, which are curved on the
detector, by tracing the slit edges, then flat field corrects and
determines a 2-D wavelength solution for each slit. For
each of the science frames, a $b$-spline sky model is fitted and
subtracted. The frames are finally combined resulting in a mean,
sky-subtracted, cosmic ray rejected 2D spectrum for each slitlet.

We found the pipeline failed for around 25\% of our data, largely
because our observational strategy deviated from that employed for 
the more routine DEEP2 survey. Our slits were long and untilted; this
caused the $b$-spline solutions to diverge. In those cases where
the sky subtraction from the automated pipeline was inadequate, we
re-reduced the slitlets using standard IRAF long-slit procedures
(as detailed in Section~\ref{subsec:lrisobs}).

\section{Redshift Distribution and Completeness Issues}
\label{sec:redshifts}

In this section we discuss the determination of spectroscopic
redshifts and present the redshift distribution for our $I-H>3$,
$H<20.5$ sample taking into account various sources of incompleteness.

A complete inventory of all our spectra is given in Figures 15, 16, and 17.

\subsection{Redshifts and Spectral Classification}

1-D spectra were extracted over a width of 1.3\arcsec. Sky
residuals due to imperfect sky subtraction were masked out and
interpolated over. In both the LRIS and DEIMOS spectra typically
around $\sim10\%$ of the array was masked out. Although the DEIMOS
resolution is much higher, we used a harsher sky threshold in the
masking as the background subtraction by the pipeline was not always
quite adequate (Section~\ref{sec:deimosobs}).
 
To facilitate redshift measurement, we compared with early and late type spectral 
templates from the Gemini Deep Deep Survey (GDDS; Abraham et al.\ 
2004\nocite{agm+04}) and a luminous red galaxy template from the Sloan 
Digital Sky Survey (SDSS; Eisenstein et al.\ 2003\nocite{ehf+03}), matching 
spectral features and the general continuum shape by visual inspection. 
Typically, identifications were based on the 4000\AA\ break, CaII H\&K (3968/3933\AA) absorption, G-band (4300\AA) 
absorption, or the [O{\scriptsize~II}]\,$\lambda\lambda$\,3726.2,3728.9\,\AA\ 
emission line doublet which was resolved in the DEIMOS data.

\begin{figure}
\psfig{figure=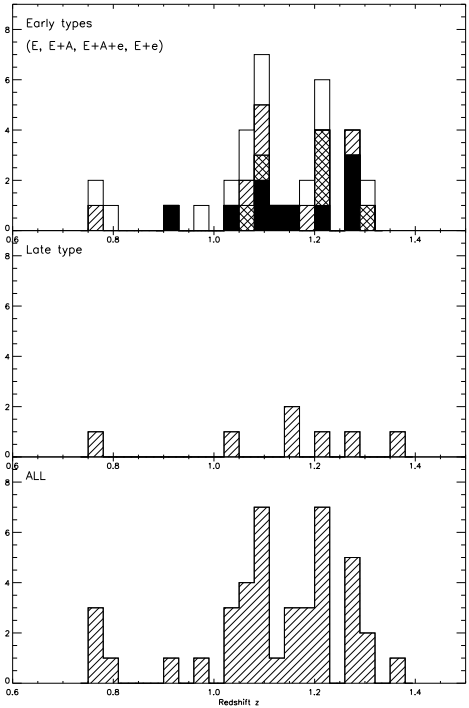,width=70mm} \\
\caption{Combined redshift distributions across the three LCIR fields for various
  spectroscopic classes. From bottom to top: all galaxies, late type, early
  types (subdivided in E, E+A (diagonal fill) / E+A+e (cross-hatched) / E+e(solid fill)).}
\label{fig:red-distrib}
\end{figure}

The redshifts for each of the 3 fields are summarised in Tables~\ref{tab:cdfs_z},
\ref{tab:ntt_z} and \ref{tab:ssa22_z}. Quality flags have been
assigned as follows: 0=fail, 1=uncertain (based on weak feature(s)), 2= probable
(based on one reliable feature), 3=secure (based on more than one certain feature) 
and -1=possible high redshift ($z>1.4$) on the basis of a strong continuum detection 
but no features (discussed below).  The redshift distribution for the whole sample is 
shown in Figure~\ref{fig:red-distrib}.  Two (5\%) of our sources are Galactic 
stars.

Of 75 objects observed in the three fields, we were able to
identify redshifts for 44 sources (including two stars).
 Of these, seven identifications ($\sim15\%$)
have the lowest quality flag (1=uncertain). Where we could not identify redshifts, it 
was usually because the continuum was too faint to detect absorption lines. 
In five cases we {\em do} detect significant continuum but fail to identify
any spectral features (quality=-1). Most likely these are higher redshift
($z>1.4$) sources where the most prominent spectral features
are redshifted beyond our spectroscopic range. In a few
cases ($\simeq$5\%), there were data acquisition or
reduction problems -- such as a source falling on top of bad pixel rows
on the detector.  In SSA22 two objects fell right at the edge of the
slit mask and were severely vignetted. In the NTT field, six sources
fell on an area of the chip which was misaligned due to one of the
alignment stars being significantly off centre (the northern most part
of the mask, Figure~\ref{fig:sky-distrib}) and the slitlets therefore
missed the galaxies. For our statistical analyses we will treat these eight (two in SSA22 and six in NTT)
sources as unobserved. Hence, our sample of observed objects is actually 67 in total, for which we obtain 44 redshifts, thereby giving $\sim66\%$ completeness.
This is comparable to or better than the success rates in other spectroscopic surveys of faint galaxies
with $R-K >5$ or $I-K >4$
(e.g. equivalent to the K20 survey work which has reached a spectroscopic completeness of 62\% to $K_s<20$, $R-K_s>5$ (Cimatti et al. 2002, 2003); see also surveys described in Yan, Thompson, and Soifer 2004; Abraham et al. 2004).
We also classified our galaxy spectra according to the prominence of 
diagnostic spectral features, as follows:
\renewcommand{\theenumi}{\roman{enumi})}
\begin{enumerate}
\item early-type (E): strong absorption features especially Ca II H\&K, no detectable 
 [OII] emission and weak or absent Balmer absorption (i.e. H$\delta$ with
 rest-frame equivalent width $<4$\AA ).
\item E+A: as early-type (i.) but with enhanced Balmer absorption (H$\delta$ with 
  rest-frame equivalent width $>4$\AA\ (see e.g. Van Dokkum \& Ellis 2003\nocite{ve03} ),
\item E+e: as early-type but with weaker Ca II H\&K and moderate to strong [O II]
emission, 
\item late-type: showing prominent
 [OII] emission ($>10\sigma$, $f\sim2.0\times10^{-18}{\rm erg~s^{-1}~cm^2}$)
 and no significant Ca II H\&K.
\item E+A+e ('mixed'): resemble the E+A class of galaxy, but also exhibit [OII] emission, i.e. 'mixed' stellar populations.
\item other: stars and AGN. There is only one AGN in our sample, which has been classified by the presence of high ionisation lines ([NeV],[NeIII]) 
\end{enumerate}

The split in spectral classifications for those sources with redshifts
is shown in Table~\ref{tab:types}.

\begin{table*}
\begin{tabular}{cccccccc}
\hline
Field & completeness &late type& early type & E+A   & E+e  &  E+A+e &other \\
\hline
CDFS &    56\%       & 0       & 45\%    &   0\% &   22\% &   22\% & 1 star ($\sim11\%$)\\
NTT  &    87\%       & 5\%     & 20\%    &  20\%  &  30\%   & 15\%   &  1 star ($\sim5\%$), 1 AGN ($\sim5\%$)\\
SSA22&    54\%       & 40\%    & 27\%    &  13\% &   13\%   &  7\%  & 0            \\
Whole Sample& 66\%  &  16\%    & 27\%    &  13.5\% & 23\%    & 13.5\% & 2 stars ($5\%$), 1 AGN ($\sim2\%$)\\
\hline
\end{tabular}
\caption{Classification of sources by field.}
\label{tab:types}
\end{table*}

\subsection{Effects of Incompleteness}

\begin{figure*}
\begin{tabular}{cc}
\psfig{figure=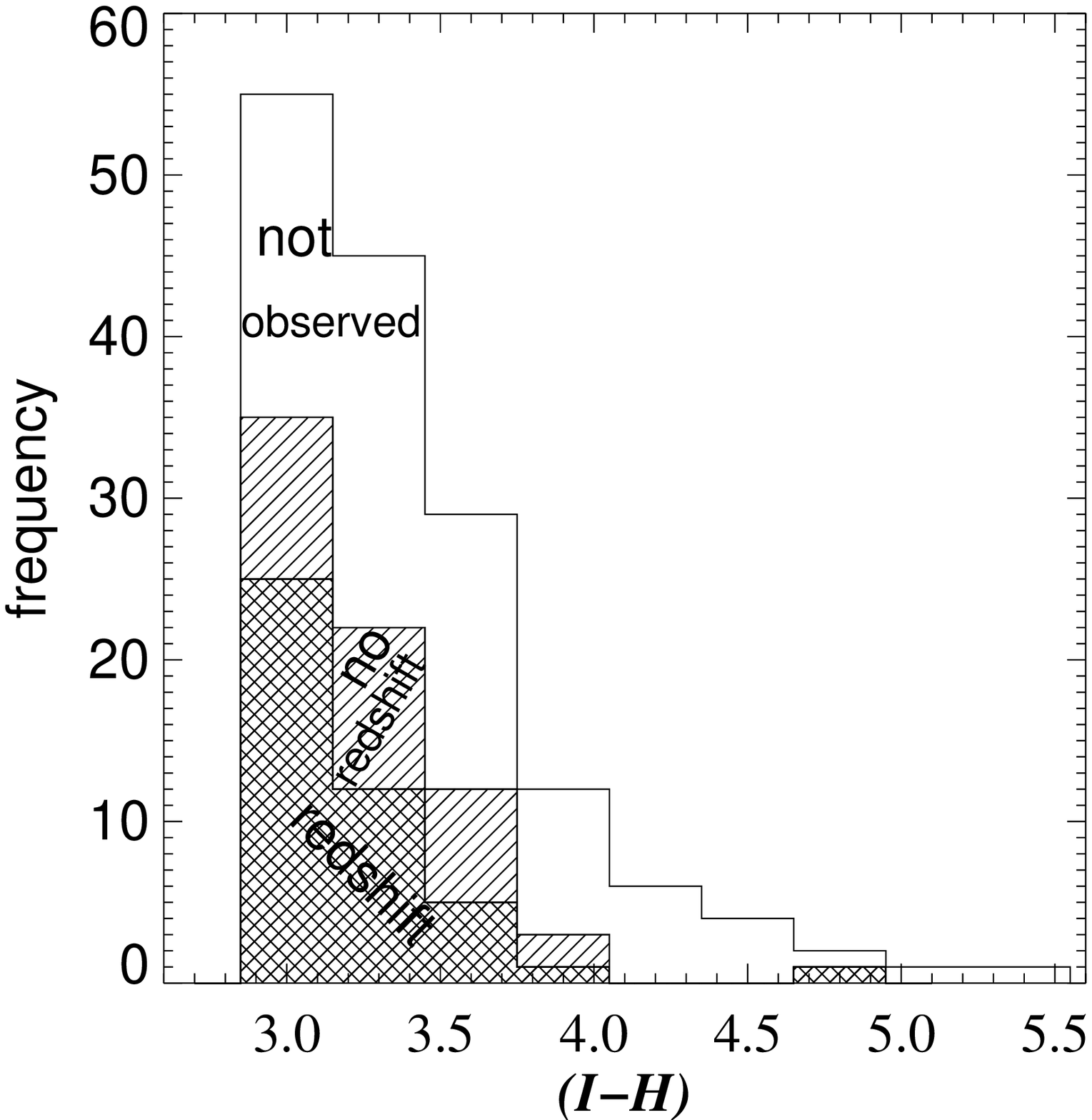,width=70mm} &
\psfig{figure=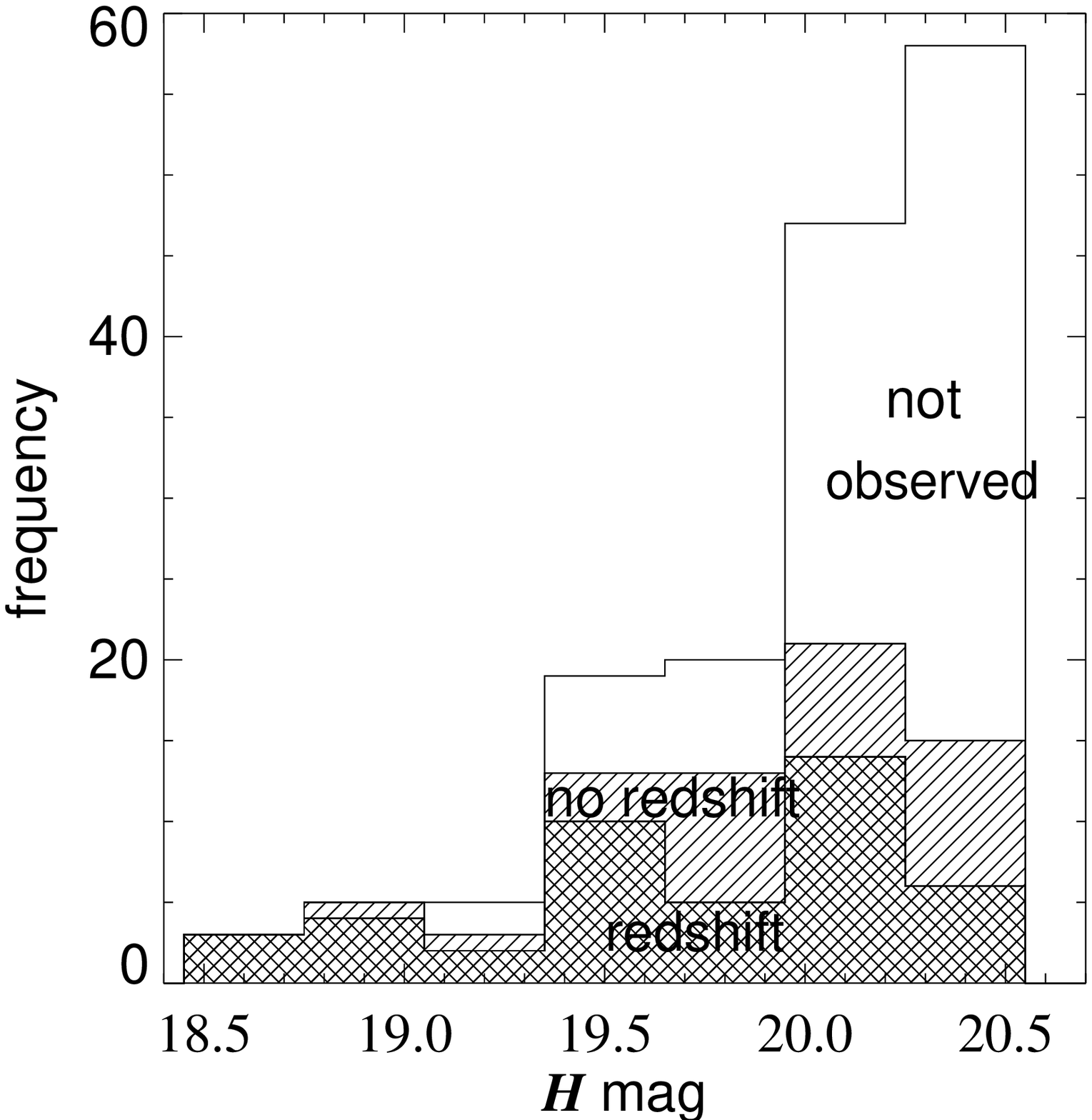,width=70mm} \\
\textbf{(a)}  & \textbf{(b)} \\
\end{tabular}
\caption{Histograms in (a) $I-H$ and (b) $H$ magnitude showing incompleteness trends in our sample. There appears to be some incompleteness towards the red ($I-H>3.5$)and faintest ($H>20$) end of the distribution, which may be accounted for by possible $z>1.4$ sources which are too red and/or faint to be detected optically.}
\label{fig:CMD}
\end{figure*}

A key question is whether the 34\% incompleteness correlates with $I-H$ colour.
For example it might be expected that sources with prominent features and large
4000 \AA\ breaks would yield redshifts more readily. Figure~\ref{fig:CMD} shows
how incompleteness correlates with colour and magnitude and also addresses
the question of the extent to which our spectroscopic targets are representative
of the overall distribution of EROs. The histograms in $I-H$ and $H$-magnitude show numbers of EROs in the detector footprints in all fields, subdivided into those for which we have identified redshifts, not identified and not observed. It is clear that the subsample of identified redshifts is a representative sample of EROs at least to $I-H\leq3.5$ and $H\leq20$, with no bias in colour or magnitude space. There is some apparent incompleteness towards the redder and fainter end of the distribution, which may possibly be populated by $z>1.4$ galaxies which are too red and faint to be detected optically.  

Of greater concern is the likelihood that some of the 34\% of unidentified 
sources lie at higher redshifts because the principal spectroscopic features
are shifted beyond our wavelength range, viz. $z>$1.4. As discussed, a
 fraction of our sample ($\sim10\%$) have quality flags
of -1 indicating a strong continuum in which features might be expected.
A key question is whether most or all of these objects lie beyond $z\simeq$1.4.

\begin{figure}
\psfig{figure=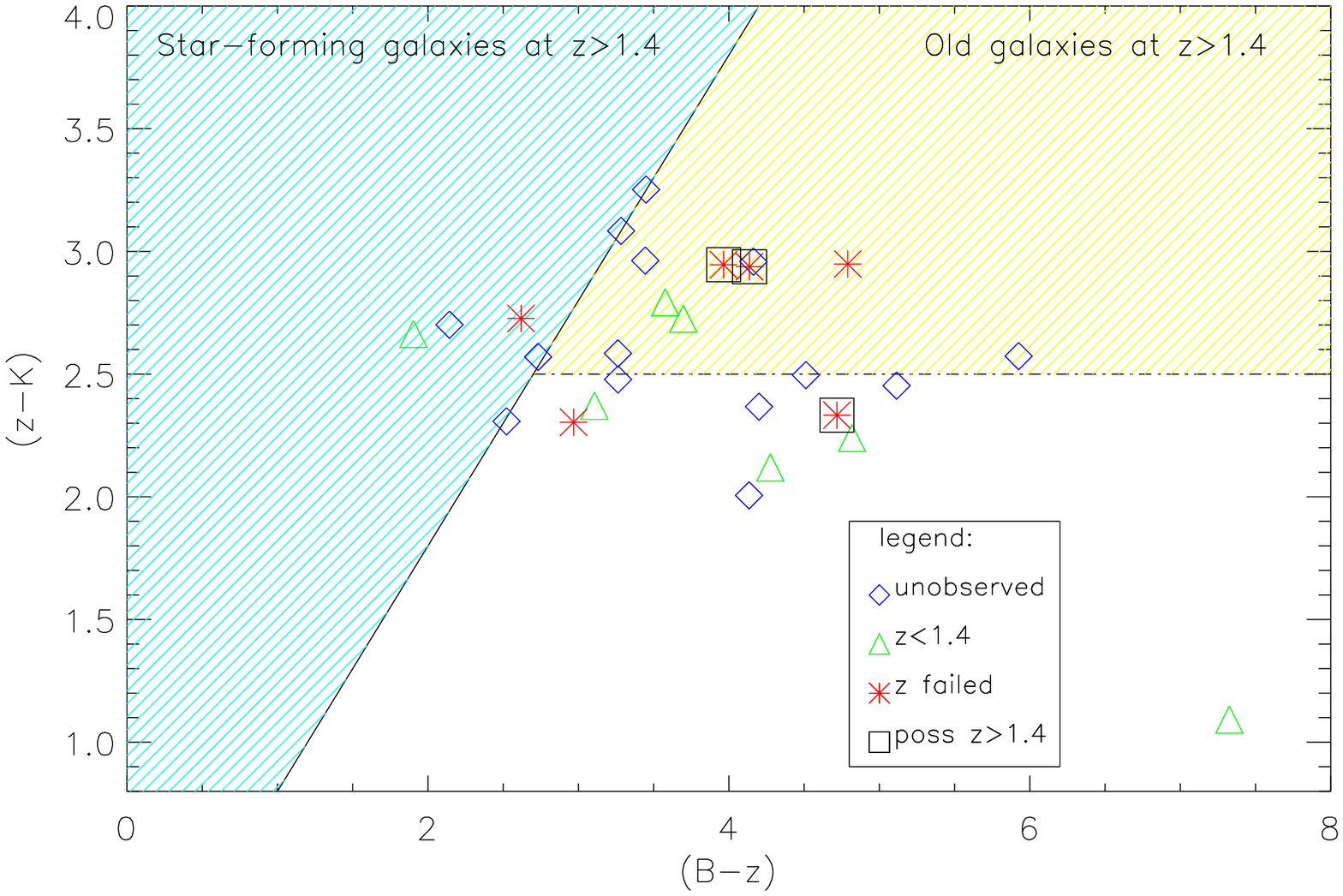,width=80mm}\\
\caption{Colour-colour $(z-K)_{AB}$ vs $(B-z)_{AB}$ diagram for galaxies in
  the CDF-S. According to Daddi et al. (2004), the region defined by $(z-K)_{AB}
  - (B-z)_{AB} \geq -0.2$ (blue shaded area) should isolate $z>1.4$
  star-forming galaxies. Conversely, the intersection of the regions
  $(z-K)_{AB} - (B-z)_{AB} \leq -0.2$ and $(z-K)_{AB} > 2.5$ should isolate
  old galaxies with $z>1.4$ (yellow shaded area). 4 out of 6 galaxies
  without redshifts (marked with an asterisk) lie in these regions
  (although the contamination from lower redshift sources is somewhat
  higher than the fraction quoted by Daddi et al. (2004). Sources flagged as
  potential $z>1.4$ sources are overlaid with boxes - 2 out of 3 fulfill
  the necessary colour criteria. }
\label{fig:BzK}
\end{figure}

\begin{figure}
\hspace{-1.0cm}
\psfig{figure=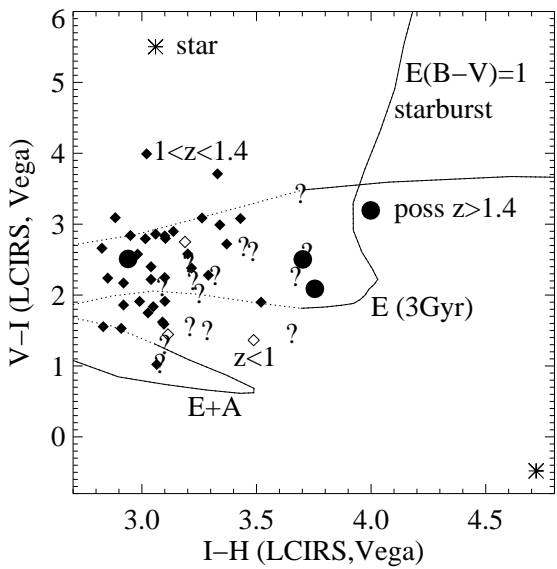,width=80mm}
\caption{$V-I$ v's $I-H$ colour-colour diagram. Only galaxies with
  spectroscopically identified redshifts with class $>1$ are shown. 
Open diamonds are $z<1$ galaxies, filled diamonds are $1<z<1.4$, and large filled circles are unidentified but believed to be at $z>1.4$ on the basis of
  featureless continuum. The remaining unknown redshifts are marked '?' and stars are marked '*'. Galaxy tracks are marked to give an indication of where $z>1.4$ galaxies might be expected to lie. The tracks are dotted at $z<1.3$ and solid at $z>1.3$.}

\label{fig:VIH}
\end{figure}

We tested for this hypothesis in the CDFS field using the photometric
redshift technique described in Daddi et al.\ (2004)\nocite{dcr+04}. These
authors have proposed that galaxies at $z>1.4$ will lie in a well-defined 
region on a $BzK$ colour-colour diagram. To facilitate the comparison we took 
F450W ($B$) and F850LP ($z$) photometry from the GOODS-South HST/ACS 
(Dickinson et al.\ 2003\nocite{dg+03}; Giavalisco et al.\ 2004\nocite{gfk+04}) and $Ks$ band data
from the ESO ISAAC survey (Vandame et al., {\it in prep}). The $Ks$ band image covers a
13 of our EROs, and of the 6 in this subsample for which we failed to derive
redshifts, four lie in the $BzK$ region for expected $z>1.4$
(Figure~\ref{fig:BzK}). However, the region is contaminated by three lower
$z$ objects, a somewhat higher fraction than that quoted by
Daddi et al.\ (2004) \nocite{dcr+04}. So while this seems to indicate
that some of our unidentified sources most likely do lie at $z>1.4$, the technique is
not definitive for a small sample. The result is strengthened however, by
the distribution of our galaxies in the $V-I$ v's $I-H$ two-colour plane (Figure~\ref{fig:VIH}),
which is also generally effective in isolating $z>1.4$ galaxies (McCarthy
2004).

\nocite{dcr+04}  

\begin{table*}
\begin{tabular}{cccccccccc}
\hline
ID    &  RA (J2000) & Dec (J2000)          & I mag & H mag & Instrument & redshift & quality & nature & comments \\
\hline
  939 &  3  32  14.98 & -27  42  25.00 & 22.45 & 18.83 &  lris & --- &  0 &        --- &          \\
 1276 &  3  32  19.14 & -27  40  40.40 & 22.00 & 18.73 &  lris &  1.1290 &  3 &     E+e &         \\
 1430 &  3  32  21.15 & -27  41   6.80 & 22.40 & 19.35 &  lris &  1.1097 &  3 &     E+A+e &         \\
 1447 &  3  32  21.24 & -27  40   9.70 & 22.55 & 19.51 &  lris &  1.0390 &  1 &     early  &      \\
 1665 &  3  32  24.16 & -27  42  11.00 & 21.90 & 18.84 &  lris &  0.0000 &  3 &      M star &     \\
 2191 &  3  32  30.51 & -27  40  30.40 & 23.46 & 20.21 &  lris & --- &  0 &        --- &          \\
 2459 &  3  32  33.87 & -27  42   4.10 & 23.48 & 20.42 &  lris &  1.2270 &  2 &      early &      \\
 2588 &  3  32  35.62 & -27  43  10.20 & 22.74 & 19.64 & deimos &  1.2210 &  3 &    E+A+e &   \\
 2668 &  3  32  36.27 & -27  42  49.50 & 23.93 & 20.18 & deimos & --- & -1 &        --- &  possible $z>1.4$  \\
 2681 &  3  32  37.18 & -27  46   8.20 & 21.67 & 18.57 & deimos &  1.0960 &  3 &    early & \\
 2800 &  3  32  39.61 & -27  47   9.00 & 23.35 & 20.14 & deimos &  1.3180 &  2 &    early & \\
 3695 &  3  32  41.64 & -27  41  52.10 & 23.09 & 19.92 & deimos & --- &  0 &        --- &  \\
 3807 &  3  32  43.92 & -27  42  32.40 & 23.67 & 19.67 & deimos & --- & -1 &        --- &  possible $z>1.4$ \\
 4071 &  3  32  42.81 & -27  43  28.90 & 23.80 & 20.10 & deimos & --- & -1 &        --- &  possible $z>1.4$ \\
 1792 &  3  32  25.77 & -27  43  47.60 & 23.48 & 20.28 & deimos & --- &  0 &        --- &  \\
 2158 &  3  32  30.33 & -27  45  23.60 & 22.54 & 19.44 & deimos &  1.2220 &  3 &     E+e &  \\
\\
\hline
I1843\footnote{Id's prefixed with an I are objects selected using $R-K>5$} &  3  32  26.28 & -27  45  36.00 & 22.78 & 19.93 & deimos &  1.2250 &  3 &        E+A &  \\
I2897 &  3  32  38.12 & -27  44  32.90 & 22.48 & 19.65 & deimos &  1.2210 &  2 &        early &  \\
I2331 &  3  32  32.97 & -27  41  16.90 & 21.66 & 18.68 & deimos &  1.0420 &  3 &        E+A+e &  \\
I3063 &  3  32  39.51 & -27  41  17.30 & 22.71 & 19.83 & deimos &  1.0394 &  3 &        early &  \\
I2876 &  3  32  38.43 & -27  40  19.40 & 22.72 & 19.90 & deimos &  1.0370 &  3 &        early &  \\
I2163 &  3  32  30.00 & -27  47  26.40 & 23.62 & 20.19 & deimos &  0.9900 &  1 &        early &  \\
\hline
\end{tabular}
\caption{CDFS Sample: Quality flags: 0=fail, 1=uncertain, 2=probable, 3=secure,  
-1=z$>1.4$,6=star. Objects prefixed with an `I' in the above table, were selected 
with $R-K>5$.}
\label{tab:cdfs_z}
\end{table*}
\begin{table*}
\begin{tabular}{ccccccccl}
\hline
ID    &  RA (J2000) & Dec (J2000)     & I mag & H mag &  $I-H$ & redshift & quality & nature \\
\hline                               
                                     
275 &   12 04 22.73  & -07 31 15.6 &  22.90 &  19.96 & 2.94 &  --- &  -1 &       poss hi-z\\
716 &   12 04 34.60  & -07 30 46.3 &  23.05 &  19.99 & 3.06 &  1.1865 &   3 &    E+A\\
1193 &  12 04 35.45  & -07 30 18.9 &  22.78 &  19.69 & 3.09 &  1.2680 &   3 &    E+e\\
1218 &  12 04 32.04  & -07 30 10.9 &  22.56 &  19.55 & 3.01 &  --- &  -1 &       poss hi-z\\
1388 &  12 04 19.58  & -07 29 54.2 &  23.59 &  20.07 & 3.52 &  1.2700 &   3 &    E+e\\
1494 &  12 04 26.70  & -07 29 46.3 &  23.13 &  19.84 & 3.29 &  1.2680 &   3 &    E+e\\
1496 &  12 04 18.98  & -07 29 45.4 &  23.45 &  20.50 & 2.95 &  1.1020 &   2 &    early\\
1962 &  12 04 24.95  & -07 29 07.4 &  23.41 &  20.21 & 3.20 &  --- &  -1 &       poss hi-z\\
2159 &  12 04 28.38  & -07 28 53.0 &  23.44 &  20.07 & 3.37 &  1.2680 &   3 &    E+A\\
3249 &  12 04 14.99  & -07 27 43.4 &  22.41 &  19.37 & 3.04 &  1.1000 &   3 &    E+A\\
3300 &  12 04 19.11  & -07 27 33.4 &  23.59 &  20.31 & 3.28 &  1.0660 &   3 &    early\\
3456 &  12 04 18.77  & -07 27 27.2 &  22.71 &  19.51 & 3.20 &  1.0700 &   3 &    E+A+e\\
3464 &  12 04 08.21  & -07 27 26.3 &  22.27 &  19.35 & 2.92 &  1.1750 &   3 &    AGN(?)\\
3472 &  12 04 18.22  & -07 27 18.1 &  23.39 &  19.96 & 3.43 &  1.0740 &   3 &    E+A\\
3720 &  12 04 16.12  & -07 27 01.2 &  23.49 &  20.15 & 3.34 &  1.0700 &   3 &    early\\
4206 &  12 04 17.36  & -07 26 26.2 &  22.69 &  19.36 & 3.33 &  1.1450 &   3 &    E+e\\
5928 &  12 04 02.05  & -07 24 18.9 &  23.06 &  20.15 & 2.91 &  1.3185 &   3 &    E+A+e\\
5962 &  12 04 07.80  & -07 24 15.9 &  23.04 &  20.12 & 2.92 &  1.1000 &   3 &    E+e\\
6052 &  12 03 57.36  & -07 24 15.9 &  23.06 &  20.14 & 2.92 &  0.9600 &   1 &    early\\
6056 &  12 04 01.39  & -07 24 09.6 &  23.00 &  20.01 & 2.99 &  1.2230 &   3 &    E+A+e\\
6322 &  12 03 56.88  & -07 23 58.9 &  22.24 &  19.20 & 3.04 &  1.1030 &   3 &    E+e\\
6696 &  12 03 55.82  & -07 23 30.1 &  22.64 &  19.27 & 3.37 &  1.2180 &   1 &    late\\
7463 &  12 03 45.06  & -07 22 32.0 &  23.71 &  18.99 & 4.72 &  0.0000 &   6 &    star\\

\hline
\end{tabular}
\caption{NTT Sample}
\label{tab:ntt_z}
\end{table*}
\begin{table*}
\begin{tabular}{cccccccccc}
\hline
ID    &  RA (J2000) & Dec (J2000)          & I mag & H mag & mask\# & redshift & quality & nature& \\
\hline
  403 & 22  17  49.01 &   0  22  14.30 & 23.46 & 20.44 &  2 & --- &  0    & ---&     \\
  527 & 22  17  41.50 &   0  21  43.60 & 23.58 & 20.29 &  2 & --- &  0    &--- &     \\
  531 & 22  17  44.84 &   0  21  42.00 & 23.58 & 20.40 &  2 & --- &  0    &---  &    \\
  713 & 22  17  40.86 &   0  20  54.50 & 23.51 & 20.42 &  2 & 1.2890 &  2 & late&    \\
  749 & 22  17  44.03 &   0  20  45.90 & 23.52 & 20.47 &  2 & --- &  0    & --- &    \\
  781 & 22  17  47.46 &   0  20  26.90 & 22.04 & 18.86 &  2 &  0.9146 & 3 &  E+e &    \\
  958 & 22  17  46.73 &   0  19  53.90 & 23.50 & 20.49 &  2 &  1.2252 & 3 &  E+A+e &   \\
 1051 & 22  17  48.87 &   0  19  26.90 & 21.80 & 18.64 &  2 &  1.2280 & 2 & early &  \\
 1267 & 22  17  49.36 &   0  18  39.60 & 23.26 & 20.19 &  2 &  1.1076 & 3 & E+A&     \\
 1535 & 22  17  35.46 &   0  17  34.60 & 23.45 & 19.81 &  1 & --- &  0    & --- &    \\
 1623 & 22  17  48.27 &   0  17  12.30 & 22.92 & 19.86 &  1 & --- &  0    & --- &    \\
 1655 & 22  17  48.59 &   0  17   5.80 & 23.39 & 19.85 &  2 &  1.1990&  1 & early &    \\
 1792 & 22  17  45.73 &   0  16  36.60 & 23.31 & 19.55 &  1 & --- &  0    & --- &    \\
 1803 & 22  17  34.62 &   0  16  36.30 & 23.37 & 20.30 &  1 & --- &  0    & --- &    \\
 1920 & 22  17  48.37 &   0  16   9.00 & 23.69 & 20.10 &  2 &  1.1420 & 1 & late&    \\
 2220 & 22  17  41.04 &   0  15   1.30 & 23.89 & 20.30 &  2 &  0.7550 & 1 & late&    \\
 2244 & 22  17  43.02 &   0  14  51.00 & 23.46 & 19.77 &  1 & --- &  0    & --- &    \\
 2260 & 22  17  32.86 &   0  14  48.10 & 22.99 & 19.50 &  1 &  0.7980 & 2 & early&    \\
 2335 & 22  17  42.08 &   0  14  30.80 & 23.37 & 19.91 &  1 & --- &  0    & --- &    \\
 2467 & 22  17  35.79 &   0  13  52.40 & 22.12 & 18.99 &  1 &  1.0320 & 3 & E+e&     \\
 2593 & 22  17  47.54 &   0  13  27.50 & 23.01 & 19.34 &  1 & --- &  0    & ---&     \\
 2638 & 22  17  41.71 &   0  13  18.00 & 23.32 & 20.21 &  1 &  0.7649 & 3 & E+A&     \\
 2767 & 22  17  39.50 &   0  10  46.50 & 22.78 & 19.56 &  1 & --- &  0    & --- &    \\
 2779 & 22  17  47.18 &   0  10  41.30 & 22.89 & 19.87 &  1 &  1.0220 & 2 & late&     \\
 2993 & 22  17  44.86 &   0  12   6.70 & 22.91 & 19.88 &  1 &  1.3695 & 2 & late&     \\
 3116 & 22  17  43.25 &   0  13   1.10 & 23.23 & 19.82 &  1 & --- &  0    & --- &    \\
 3253 & 22  17  32.34 &   0  12  52.40 & 23.25 & 20.15 &  1 &  1.1530 & 2 & late&    \\
 3378 & 22  17  33.67 &   0  13   5.30 & 23.38 & 19.54 &  1 &  0.7620 & 1 & early?&  \\
\hline
\end{tabular}
\caption{SSA22 Sample}
\label{tab:ssa22_z}
\end{table*}

\subsection{Comparison with Earlier Work}

There have been several redshift surveys of faint objects in our target fields, and 
consequently there is some small overlap ($\simeq$20\%) with our ERO sample. As a 
check on our identifications, we matched our catalogue by coordinates (to within 1
arcsec), with the GDDS sample for the SSA22 field (Abraham et al.\ 2004) and 
with the FORS2 spectroscopic sample of Vanzella et al.\ (2004)
\nocite{van04} and the K20 survey\footnote{www.arcetri.astro.it/$\sim$k20} (Cimatti
et al.\ 2002) for the CDFS field. 
For the FORS2 sample (Table~\ref{tab:fors2}), we find good agreement with the five 
sources in common. Four have high confidence quality flags in both samples and agree 
to within $\Delta z=0.001$. The redshift for the remaining source is unidentified 
in both surveys. Of the four sources overlapping with the K20 survey (Table~\ref{tab:k20}), three of the redshifts are in agreement, but one which we have flagged as low confidence at z=0.99 is discrepant with the higher z=1.553 K20 estimate.  

\begin{table*}
\begin{tabular}{llllll}
\hline
ID   &  redshift & I mag & FORS2 ID &      FORS2 redshift & FORS2 I mag \\
\hline
1276  &  1.129 & 22.00 & J033219.15-274040.2 & 1.128  & 22.25 \\
2681  &  1.096 & 21.67 & J033237.19-274608.1 & 1.096 & 21.98 \\
2800  &  1.318 & 23.35 & J033239.64-274709.1 & 1.317  & 23.69 \\
2158  &  1.222 & 22.54 & J033230.34-274523.6 & 1.223  & 22.89 \\
1792  &  ---   & 23.48 & J033225.76-274347.0 &  ---   & 24.13 \\
\hline
\end{tabular}
\caption{Redshift comparison with the FORS2 survey in the CDFS field.}
\label{tab:fors2}
\end{table*}

\begin{table*}
\begin{tabular}{llllll}
\hline
 ID    &    redshift  & quality flag &  K20 id &     K20 redshift\\
 2681  &    1.09600   &3            &  CDFS\_00633 &   1.09600 \\   
 2158  &    1.22200   &3            &  CDFS\_00507 &   1.22300   \\ 
 I1843 &    1.22500   &3            &  CDFS\_00547 &   1.22200    \\
 I2163 &    0.990000  &1            &  CDFS\_00139 &   1.55300    \\
\hline
\end{tabular}
\caption{Redshift comparison with the K20 survey in the CDFS field.}
\label{tab:k20}
\end{table*}

Matches with the GDDS sample are shown in Table~\ref{tab:gdds}. Of the three overlapping sources for which we have confident identifications, one agrees and two disagree.
The GDDS confidence rating for one of these failures is $<50\%$. In three cases 
where our redshift identification is insecure (class 1) and in two cases where we 
failed, GDDS find higher redshifts ($z>1.395$) with a high confidence rating,
supporting our hypothesis that the redshifts which we failed to get
 potentially lie at higher-$z$.

 \begin{table*}
\begin{tabular}{llllllll}
\hline
ID &     redshift & quality flag & GDDS ID & GDDS redshift & GDDS I mag & GDDS confidence \\  
\hline
  3253  &  1.153 & 2 & 23.25  & SA22-0062 &   1.154 &  23.17 & [O{\scriptsize~II}]\\
  2260  &  0.798 & 2 & 22.99 &  SA22-0448 &   1.202 &  22.80  &$<$50\%\\
  2467  &  1.032 & 3 & 22.12 &  SA22-2548 &   1.022 &  21.96 &certain\\

  1655  &  1.199 & 1 & 23.39 &  SA22-0674 &   1.493 &  23.19 &$>$75\%\\
  1920  &  1.142 & 1 & 23.69 &  SA22-1983 &   1.488 &  23.68 &95\%\\
  3378  &  0.762 & 1 & 23.38 &  SA22-0107 &   1.448 &  23.20 &single emission line\\
 
  2335  & --- & 0 & 23.37 &  SA22-0398  &  1.395 &  23.16 &95\%\\
  2593  & --- & 0 & 23.01 &  SA22-0189  &  1.490 &  22.80 &95\%\\
  1535  & --- & 0 & 23.45 &  SA22-0948  &  1.396 &  23.24 &none/best guess\\
  1792  & --- & 0 & 23.31 &  SA22-0721  &  1.483 &  23.14 &none/best guess\\
  1803  & --- & 0 & 23.37 &  SA22-0717  &  2.060 &  23.10 &none/best guess\\
\hline
\end{tabular}
\caption{Redshift comparison with the GDDS survey in the SSA22 field. }
\label{tab:gdds}
\end{table*}

\section{Analysis}
\label{sec:discussion}

We now turn to addressing the key objectives for the survey set out in 
$\S$\ref{sec:intro}. Summarizing the situation, using the photometric 
criteria $H<20.5$ and $I-H>3.0$, we have drawn spectroscopic targets 
from 252 EROs in three LCIR survey fields, of which 159 fall 
within the field of view of the slit masks employed. We successfully 
observed a sub-sample of 67 objects drawn randomly from the photometric
sample. Spectroscopic identifications were secured for 44 sources (including 
two stars), a completeness fraction of 66\%. We have demonstrated that
this large spectroscopic sample is representative of the ERO population
in the LCIR survey. The spectra are generally of good quality with higher 
dispersion than in previous studies (e.g. K20, GDDS).

\subsection{Redshift Distribution}
\label{subsec:split}

Our redshift distribution ($N(z)$, Figure~\ref{fig:red-distrib}) ranges
between 0.755 and 1.4 (with some sources potentially at $z>1.4$), and therefore demonstrably supports the 
hypothesis that an $I-H>3$ colour criterion is effective in selecting 
galaxies at $z \gtrsim 1$ (McCarthy et al.\ 2001). 

\begin{figure}
\psfig{figure=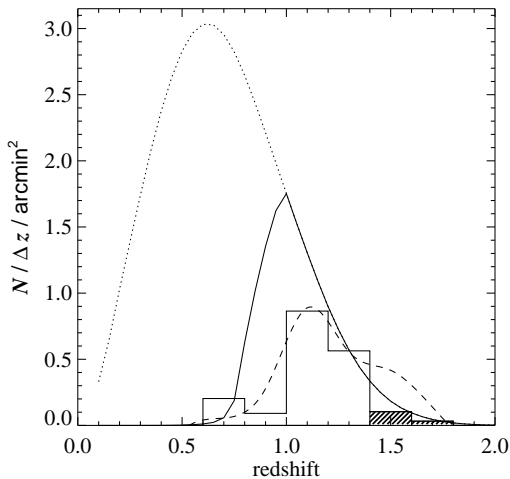,width=70mm}\\
\caption{Observed redshift distribution for 3 LCIRS fields, corrected for 
incompleteness (histogram - with shaded region possible $z>1.4$ galaxies), compared with that estimated photometrically by 
Firth et al. (2002; dashed curve) in a fourth field. Overplotted is a
predicted N(z) if EROs produced all ellipticals today via passive
luminosity evolution.  The local
elliptical population has been reverse evolved (assuming all are 5Gyr old),
using the luminosity function derived by Nakamura et al (2003) for E/S0 galaxies in the SDSS, which has $\alpha=-0.83$ (slope), $\phi$*=0.0047*(0.7)$^3$  per Mpc$^3$,
M*= -20.75+5.*alog10(0.7)) The solid curve is the backwards-PLE
model including our $I-H>3.0$ colour cut, and dotted curve has
no colour cut on PLE).
}
\label{fig:NZcomp}
\end{figure}

Figure~\ref{fig:NZcomp} compares the observed distribution and that
predicted for the larger LCIRS sample using photometric redshifts 
based on $UBVRIH$ photometry in the independent LCIRS field HDF-S 
(Firth et al.\ 2002). From the latter distribution, Firth et al.\ 
concluded that popular semi-analytic models provided a reasonable
fit to the overall $N(z)$ but under-predicted the abundance. 

\nocite{nfy+03}    

The redshift distribution in our
spectroscopic survey varies from field to field with clear structures
in each, suggesting significant cosmic variance and clustering among
the population. Furthermore, different mixtures of spectral types are
seen in the various fields (Table~\ref{tab:types}). The simplest
conclusion consistent with this disparity is that the dominant component
of the ERO population is made of luminous spheroidals which are
strongly clustered. We will return to analyzing the variance seen
across the three fields when we measure the overall spatial clustering
in our sample in $\S$~\ref{subsec:cluster}.

Although our spectroscopic redshift distribution is
lacking the high$-z$ tail seen in the photometric distribution - due to our
incompleteness in the `spectroscopic desert' -- or $z>1.4$ regime,
we find an overall abundance in agreement with that found by
Firth et al. in the HDF-S, i.e. significantly more than is predicted by
semi-analytic models. 

\subsection{$I-H$ versus $R-K$ Colour Selection}

Our $I-H>3$ selection criterion is different from the frequently-used
$R-K>5$ (Cimatti et al.\ 2002) or $R-K>6$ (Hu \& Ridgway 1994) cut and
thus it is interesting to consider whether the greater prominence of
early-type spectra (and the larger associated cosmic variance) is due,
in part, to this difference. Our fraction of actively star-forming galaxies 
($\sim16\%$) is much less than that found (50\%) in the $R-K$ selected 
K20 sample (Cimatti et al. 2002) or the bright K-sample of
Yan et al. (2004)\nocite{yts04} . 

We also might reasonably expect the shorter wavelength baseline of $I-H$ (compared
to $R-K$) to provide an increased
sensitivity to strong age-dependent continuum breaks as opposed to
the broader wavelength signatures arising from reddening. Unfortunately,
given the different nature and classification methods for the various 
spectroscopic datasets, it is unclear whether the different fractions of 
early-type galaxies arises entirely from the colour selection.  Part of
the difference may arise from the increased signal to noise in our data.
For example, although a further 40\% of our galaxies reveal modest 
[O{\scriptsize~II}] emission, our spectra enable us to clearly demonstrate 
an underlying old stellar population, whereas such galaxy types may previously have been classified as dust-reddened starbursts. Figure 1 also demonstrates that E+A galaxies, such as appear in our sample (for example, with 5\% burst mass in the last 100~Myr) would not be expected in a selection based on $R-K>5$ colour cut.   

\subsection{Spectroscopic Properties of the ERO Population}

We now turn to the important question of the nature of the ERO population
as revealed by the Keck spectra. Thus far we have categorized each
spectrum according to its appearance, deferring any astrophysical
interpretation of the various classifications. Our goal is to address
the relevance of the E, E+A, E+e, E+A+e and late types introduced in $\S$4.1
as well as the homogeneity of the overall ERO population. We stress
that our analysis in this section will be illustrative given that the spectroscopic
data alone cannot separate uniquely many of the key variables e.g. the effects 
of age, metallicity and complex star formation histories (c.f. McCarthy et al.
2004\nocite{mcc04}).

\begin{figure}
\begin{tabular}{c}
\psfig{figure=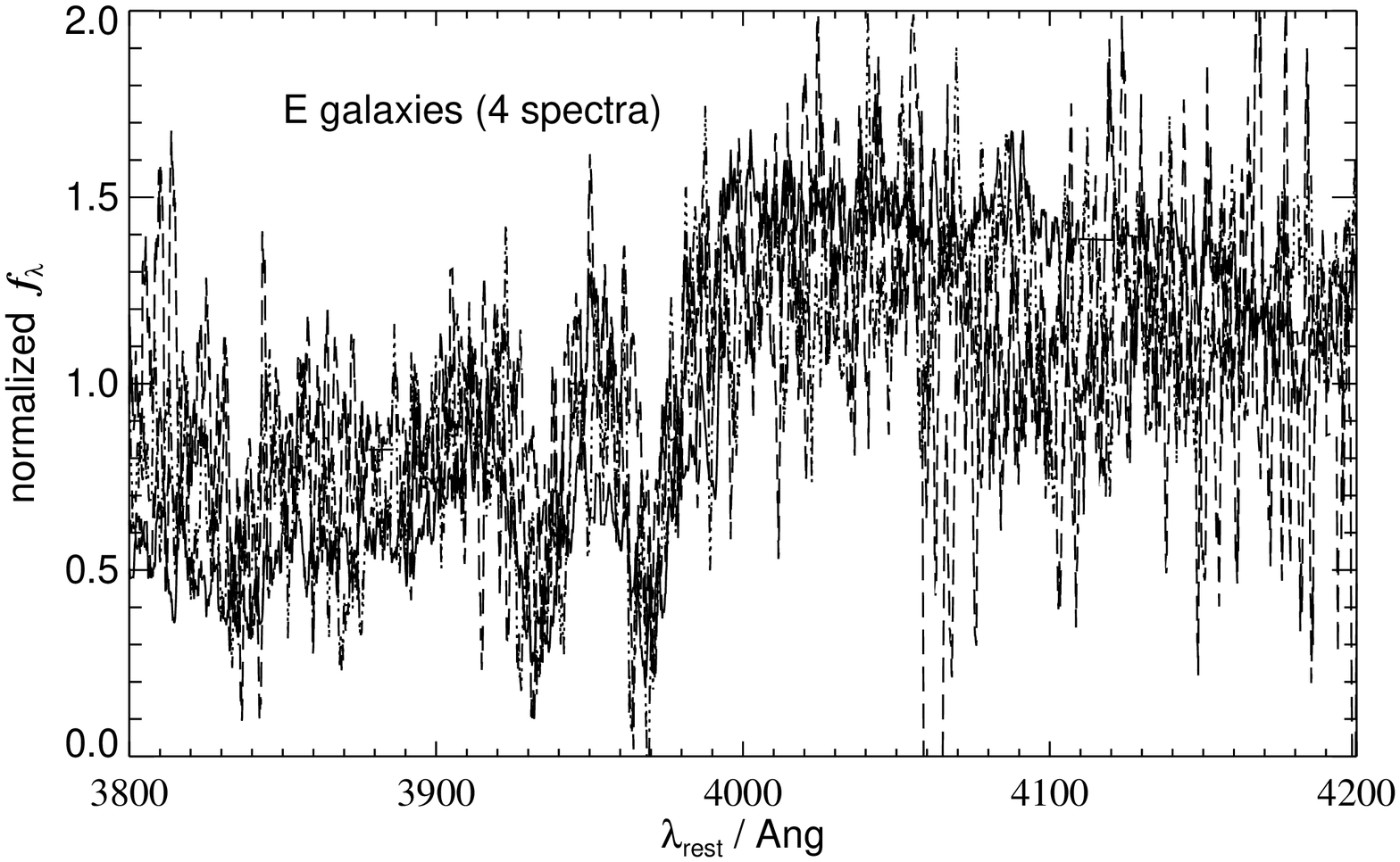,width=70mm} \\
\psfig{figure=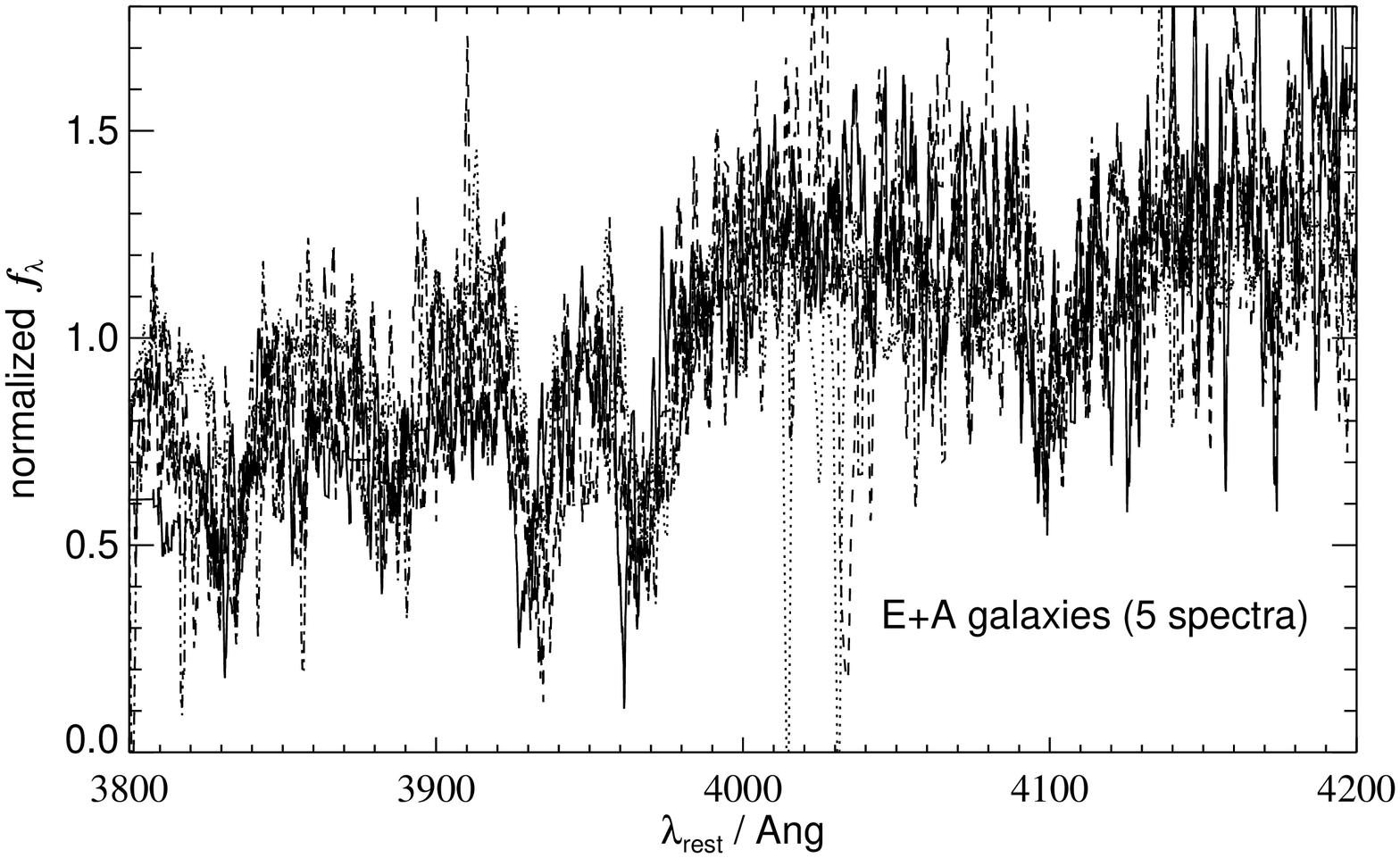,width=70mm} \\
\end{tabular}
\caption{Noise spectrum for the E (top) and E+A (bottom) composite spectra, composed of the 
  individual spectra normalised to the same continuum flux and
  overlaid. The standard deviation in this spectrum is then the error on
  the composite spectrum. }
\label{fig:noise}
\end{figure}

\begin{figure*}
\psfig{figure=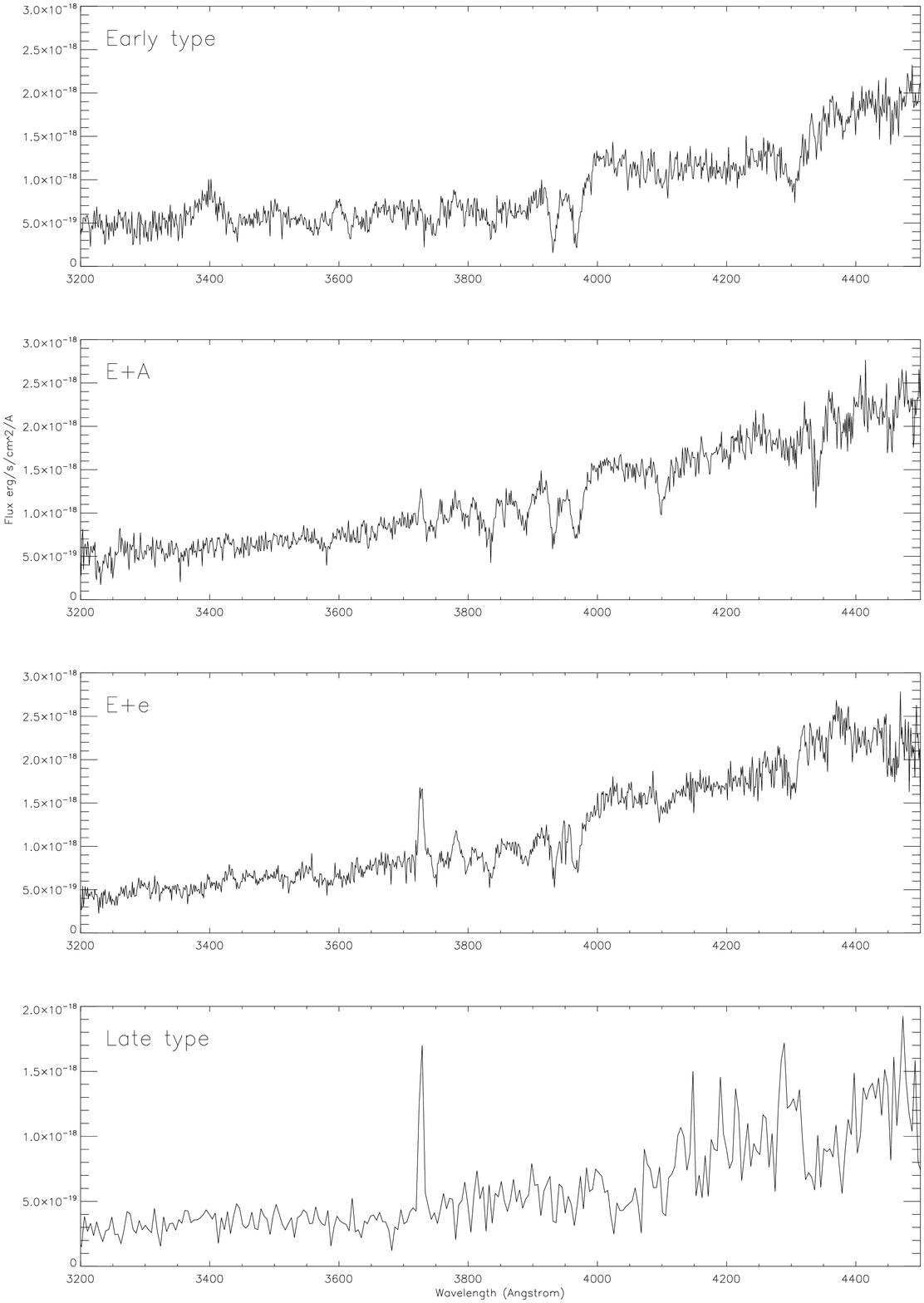,width=160mm}
\\
\caption{Stacked spectra for categories E, E+A, E+e and late type as defined in
$\S$4.1. The E, E+A and E+e spectra were drawn from the superior resolution DEIMOS
data whereas the late-type spectra were restricted to the LRIS sample. Only 
spectra with quality flags 2 or 3 were included. }
\label{fig:stacked}
\end{figure*}

We begin by stacking the spectra according to our basic classification scheme.
Figure~\ref{fig:stacked} shows the result of this coaddition for the
higher quality spectra selected with quality flags 2 or 3. In practice, given 
we used both DEIMOS and LRIS spectrographs, we restricted the coaddition of 
the E, E+A and E+e spectra, for which absorption line measures are particularly
helpful, to those taken with the superior resolution DEIMOS, rebinning to a final
dispersion of 3.5\AA pixel$^{-1}$. In the case of the late-type spectra,
as most examples occur in the LRIS sample, we restricted the coaddition
to those taken with LRIS rebinned to 13.6\AA pixel$^{-1}$. We
find the error on each composite spectrum to be the deviation in the spectra when each is normalised
to the same relative flux in the wavelength range $3800-4200\AA$. The noise spectra for
the E and E+A galaxies are shown in Figure~\ref{fig:noise}, we use this to
deduce the error on diagnostics such as the D4000 break and line equivalent
widths.

We now briefly discuss the coadded spectra in turn:

\subsubsection{Pure early type (E)}
\label{sec:E}
About $30\%$ of our sample are in the E category and our luminosity-weighted
mean spectrum has a 4000\AA\ break $D_{4000} \sim1.7$\footnote{We define the index 
$D_{4000}= f(3750-3950\AA)/f(4050-4250\AA)$ (\nocite{bru83}(Bruzual 1983)}.
 Comparing this with the hypothesis of a simple stellar population following
a single burst in the context of Bruzual \& Charlot (2003)\nocite{bc03} models we find an
average age of $\sim 2$\,Gyr at the mean redshift of $\overline{z}$=1.2, assuming 
solar metallicity (Figure~\ref{fig:D4000models}a). This implies a formation epoch 
of $z>2$ for this subset of the ERO population. This result agrees with
McCarthy et al. (2004) who find a $<z_f>$=2.4 for 20 red galaxies with
$z>1.3$ in the GDDS, and deduce an early and rapid formation for a
substantial fraction of these. Furthermore, Cimatti et al. (2002,2003,2004) deduce formation redshifts between 2--3.4 for similar objects found in the K20 survey.

\begin{figure*}
\begin{tabular}{c}
\psfig{figure=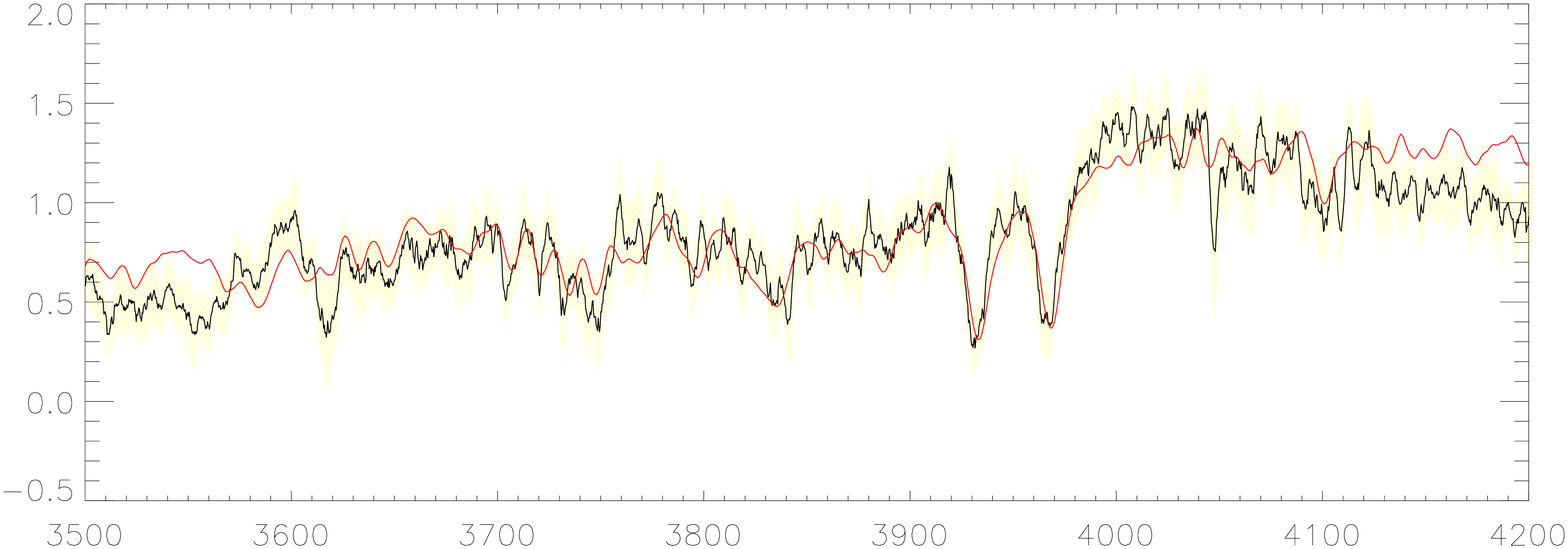,width=160mm}

\\
{\bf (a)}\\
\psfig{figure=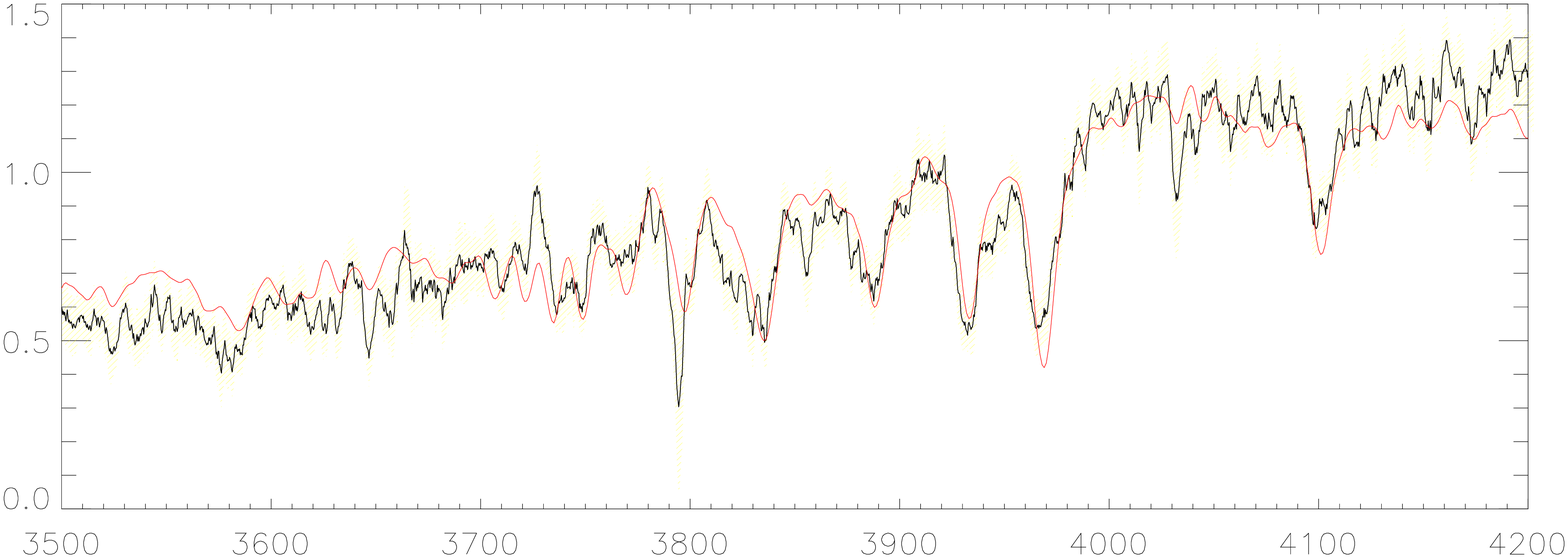,width=160mm} \\

{\bf (b)}\\
\end{tabular}
\caption{(a) Pure early type composite spectrum compared to a Bruzual and
  Charlot (2003) model spectrum (dashed line) of a 3.5Gyr population and (b)
  the E+A spectrum is well fit by an underlying  2Gyr population, with a
  100Myr old secondary burst of 2\% by mass. Both assume a Salpeter IMF and
  solar metallicity. The shaded regions show the errors on the composite
  spectra, which are the standard
  deviation of the flux at each wavelength in the individual spectra (scaled to the same
  relative luminosity). }
\label{fig:bc_spec}
\end{figure*}

\begin{figure*}
\begin{tabular}{cc}
\psfig{figure=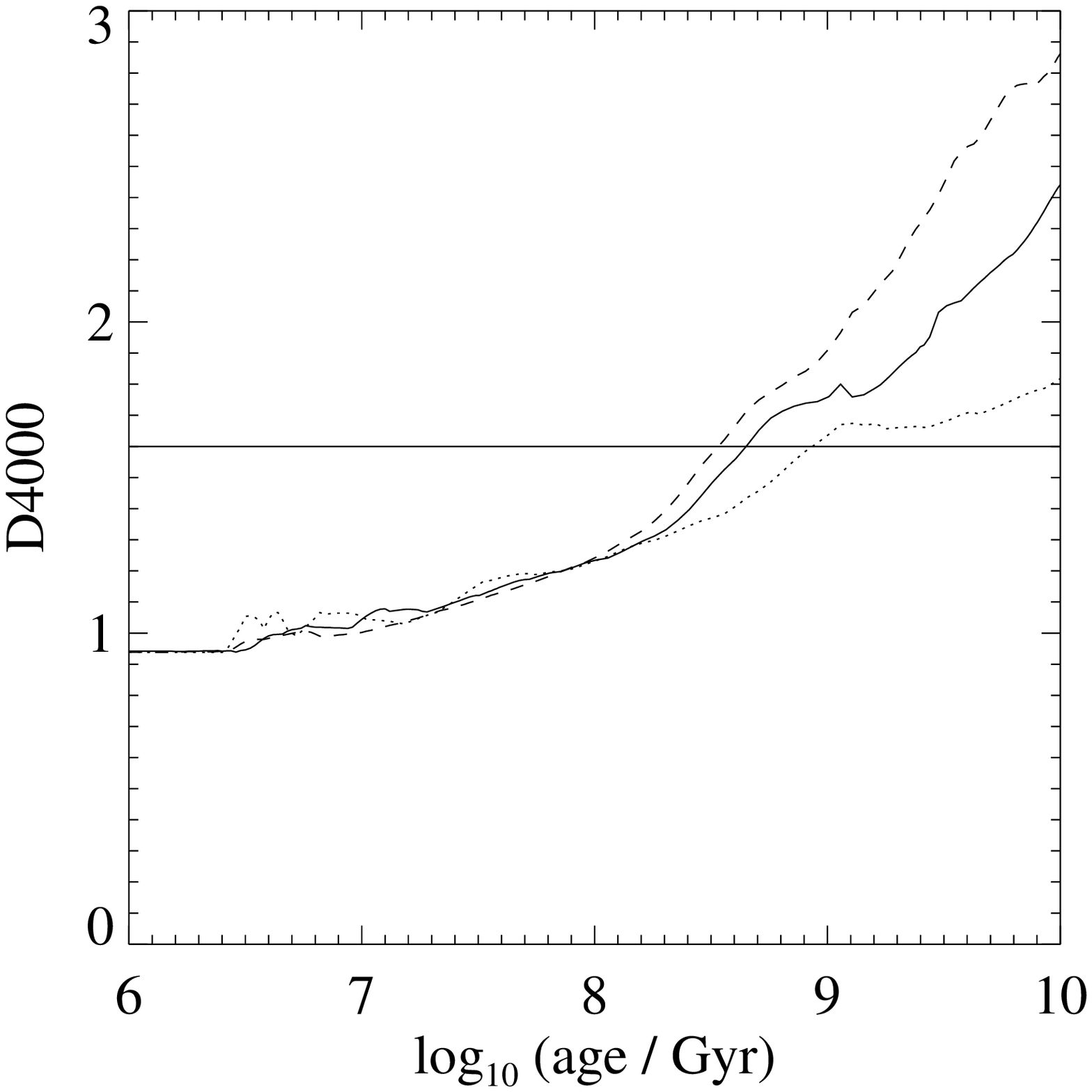,width=60mm} &
\psfig{figure=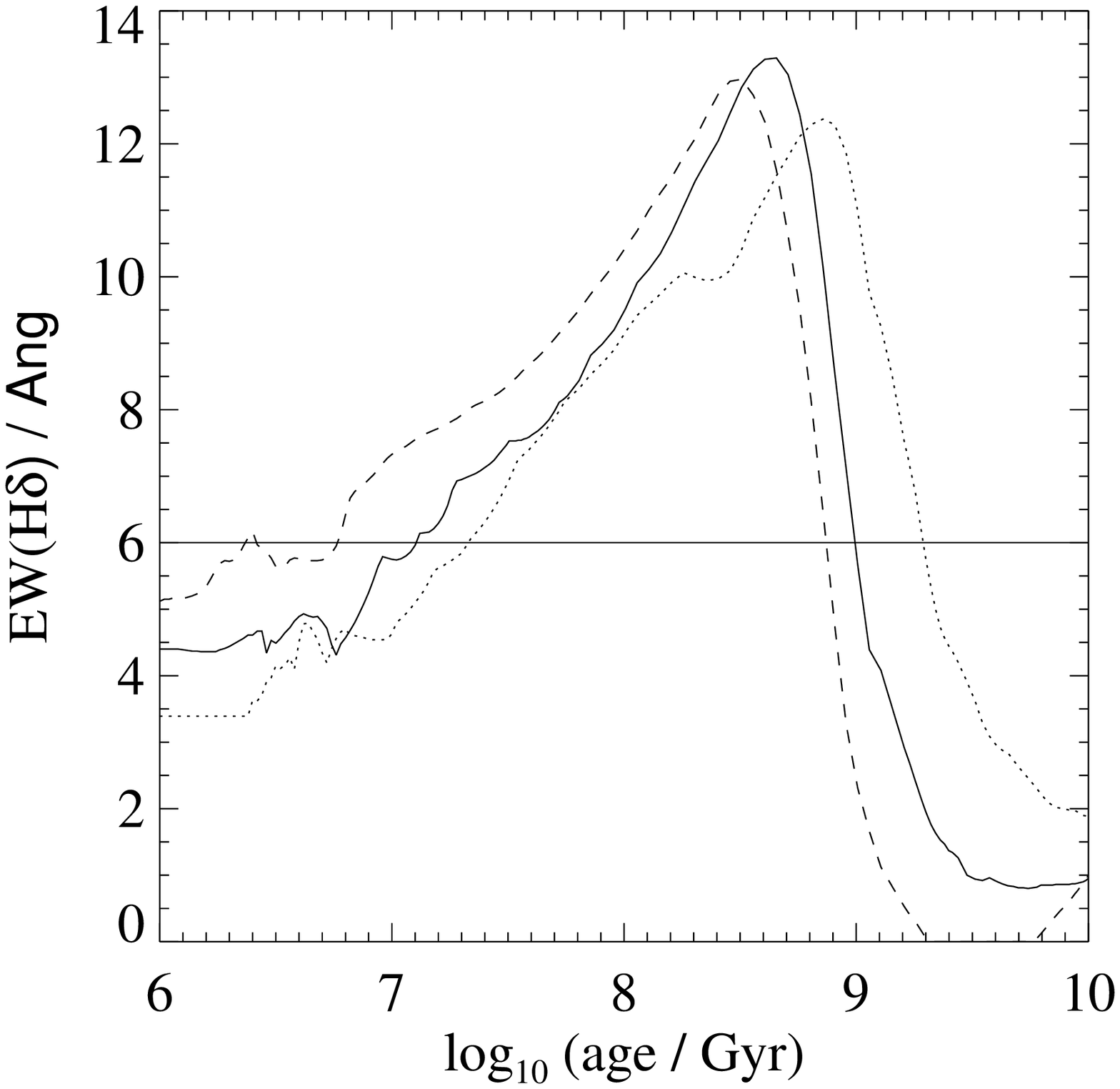,width=60mm}\\
{\bf (a)} & {\bf (b) }\\
\end{tabular}
\caption{(a) Evolution of $D_{4000}$ according to a single-burst Bruzual \& Charlot (2003) 
model of solar metallicity (solid line), 0.2$\times$solar (dotted line) and 2.5$\times$ solar (dashed line). Comparison with the mean early-type spectrum implies an 
age of $\sim1.6$ Gyr at the mean sample redshift $z\simeq$1.2. (b) Evolution of EW(H$\delta$) for a single-burst Bruzual \& Charlot (2003) 
model of solar metallicity (solid line), 0.2$\times$solar (dotted line) and 2.5$\times$ solar (dashed line).}
\label{fig:D4000models}
\end{figure*}

\subsubsection{Intermediate age (E+A)}
\label{sec:E+A}

A significant fraction of our sample have strong Balmer absorption lines
in addition to prominent 4000\AA\ breaks and are indicative of secondary
star-formation. These galaxies have traditionally been referred to as E+A
galaxies (also known as 'K+A'; Dressler et al. 1999 \nocite{dsp+99} ; see
also Blake et al. 2004\nocite{bpc+04} and refs. therein). Although $\sim30\%$ of our absorption line sample fall in this 
category, the fraction shows considerable field-to-field dispersion (0-50\% across 
the 3 LCIRS fields). Nonetheless, the rate of occurrence is much higher than in 
nearby field galaxies.

Zabludoff et al. (1996)\nocite{zzl+96} find 0.6\% of all galaxies at $z\sim0$ in the Las
Campanas Redshift Survey to be E+A (as selected by
EW(H$\delta$)$>4.5$\AA). Given that EROs make up about $\sim10\%$ of the
overall galaxy population  at our limiting magnitude, we find more than
twice as many E+As at $z\sim1$. With a stricter criterion,
(EW(H$\delta)>5\AA$) Goto (2005)\nocite{goto05} finds a local E+A incidence of
$\sim0.1\%$ in the Sloan Digital Sky Survey (SDSS; York et al. 2000\nocite{yaa+00}). Bearing in mind that our selection function for EROs is picking out only E+A galaxies which are spheroidal in origin (locally, E+As are found to also have late-type spiral progenitors, see e.g. Blake et al. 2004, Zabludoff et al. 1996), this is reasonably strong evidence for evolution in the number density of such systems from $z\sim1-2$ to $z\sim0$.

The composite spectrum (Figure~\ref{fig:stacked} has an average
equivalent width $W_{H\delta}$ = 6\AA.  Figure~\ref{fig:D4000models}b shows the evolution with age of the H$\delta$ equivalent width for sub-solar, solar and super-solar metallicities. Combining with the $D_{4000}$
index ($\S$~\ref{sec:E},Figure~\ref{fig:D4000models}) we find the spectrum can be readily interpreted in terms of
an underlying population as old as that of the mean E type spectrum discussed
above, with an additional secondary component (5-10\%) which formed 
between 30 Myr and 1 Gyr prior to the epoch of observation. This is consistent with examples of star formation activity found in $z\sim1$ spheroidals by Treu (2004 and references therein\nocite{treu04}).

In summary, the simplest hypothesis is that the E+A galaxies have a similar 
origin to the E-type galaxies reflecting that component of the population
which recently underwent a secondary burst, possibly associated with
a merger. It is possible that the K-bright star forming objects at $z\sim2$
seen in  GOODS and K20 could be the same objects we are seeing as E+As at
$z\sim1$, in their post-burst phase.  

\subsubsection{Active early type (E+e)}

As earlier studies in clusters have shown (e.g. Couch \& Sharples 1987\nocite{cs87},
Barger et al 1996\nocite{bae+96}), for established stellar systems undergoing secondary 
activity, a subset are likely to show the mixed signals of an old stellar 
population and emission lines during the active phase. In the
case of merger driven evolution, unless the accretion is purely stellar
such systems will show diluted evidence of an old stellar population and 
[O II] emission. Depending on the duty-cycle, the Balmer absorption 
lines will be partially or wholly filled by gaseous emission. In fact, although Balmer absorption lines are undetected amongst each indivual spectrum in this class, they are weakly visible in the composite spectrum (Figure~\ref{fig:stacked}). 

Given our explanation for the E+A sources, it is natural to adopt the
above explanation for many of the E+e spectra whose line properties
are understandably more diverse.  
\\

\begin{figure*}
\psfig{figure=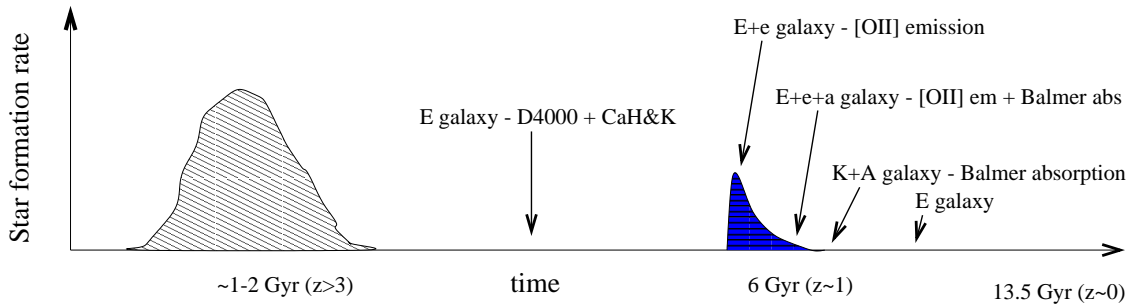,width=150mm}
\\
\caption{Diagram showing a cycle of star formation activity which would explain the observed spectra of the E, E+e, E+A and E+A with [OII]. The E type galaxies are formed in an initial discrete burst of star formation, some time later a secondary burst gives rise to the [OII] emission seen in the E+e spectra and subsequently Balmer aborption lines once star formation is truncated.}
\label{fig:SFRhist}
\end{figure*}

In summary, the dominant component of our ERO sample consists of old galaxies seen
in various stages of intermittent, but minor, activity. Figure~\ref{fig:SFRhist} shows an exemplative star formation cycle which may give rise to the various stages of evolution we observe in our sample. The E type spectra are old galaxies are formed at high redshift, perhaps in an initial discrete burst of star formation. Some time later a secondary burst is triggered, giving rise to the [OII] emission seen in the E+e spectra, and subsequently Balmer aborption lines become visible once star formation has been truncated, in the E+A class. The mixed populations seen, which are predominantly E+A with [OII] emission, represent some intermediate stage between E+e and E+A. Barger et al. (1996) successfully explain the numbers of systems similar to those discussed above in clusters at low redshift (z=0.31) using models  involving secondary bursts of star formation over the 2 Gyr prior to the epoch of observation. 
Treu et al. (2005)\nocite{treu05} have recently demonstrated that the star
formation history of field spheroidals is heavily dependant on their
mass. If galaxy formation is governed by a 'down-sizing' mechanism
(Cowie et al. 1996)\nocite{cshc96}, i.e. growth through rapid star-formation trends smoothly to
less massive galaxies with decreasing redshift, then in a large sample we might expect the E, E+A, E+A+e and E+e galaxies to be progressively ordered in decreasing luminosity (e.g. in the rest-frame $I-$band).
We examined our data, unsuccessfully, for evidence of such a trend. It may be that all of these galaxies formed the bulk of their mass at high-z and that subsequently environment plays a bigger role, with mergers or interactions triggering more recent complex star formation histories.

\subsubsection{Late Type}

A small fraction ($\sim16\%$) of our sample are classed as late type on
the basis of prominent [O{\scriptsize~II}] emission together with 
weak or non-existent Ca II H\&K absorption. In these systems, the luminosity
weighted stellar population is clearly young and star-forming.

In the co-added LRIS spectrum of 4 late-type galaxies in SSA22, we find 
an average line flux of $1\times10^{-17}{\rm erg s^{-1} cm^{-2}}$. At the 
average redshift z=1.2, this equates to a fairly modest star formation
rate (SFR) of $1.1\pm0.3$ \msol yr$^{-1}$ using Kennicutt's (1998) conversion:

$SFR(\msol yr^{-1})=(1.4\pm0.4)\times10^{-41}L([OII])(erg s^{-1})$
  
However this may be a lower limit as it is uncorrected for 
extinction which presumably is responsible for the red $I-H$ colour. 
As an illustration, a `pure' dusty starburst (with no significant
flux contribution from an underlying older population) our $I-H>3$ 
selection criterion requires $E(B-V)\gtrsim 1$ (Figure~\ref{fig:tracks}).
In the simplest case where dust screens the gas and stars by equal
amounts, this would imply a correction to the SFR as large as $\gtrsim 50$
($\sim 100$\,\msol yr$^{-1}$). in practice, of course, there are
numerous uncertainties in deriving star formation rates in this
manner. The recent Spitzer results of Yan et al. (2004) imply high
extinction for the late type red galaxies and typical star formation rates
$50-170 {\rm \msol\, yr{^-1}}$.  \nocite{ycf+04}

\subsection{Abundance of Galaxies with Established Stellar Populations}
\label{subsec:numdens}

The foregoing discussion has revealed that the bulk of our $I-H>$3 sample,
specifically most of those in the dominant spectral categories E, E+A and E+e, 
contain well-established stellar populations (defined broadly as those where
the bulk of the star formation occurred prior to $z$=2). It is thus interesting
to make a rough comparison of the luminosity density in this population with 
that in the population of present day spheroidals.

With a surface area of 200~arcmin$^{-2}$ across our 3 LCIRS fields, and assuming
a uniform redshift-dependent selection function across the interval 1.0$<z<$1.35
(in which 90\% of our objects lie) we effectively survey a co-moving volume of 
$\sim1.70\times10^4$Mpc$^3$. The average space density of EROs is thus
$\phi\approx1\times10^{-3}$Mpc$^{-3}$ of which we estimate $\sim75\%$ contain an established
stellar population. At the median redshift $z$=1.2, our $H<$20.5 magnitude
limit corresponds to an absolute rest-frame I-band luminosity of M$_I=$-21.85,
equivalent in local terms (Blanton et al. 2003 \nocite{bhh+03}) to $\simeq$1.25L$^{\ast}$. 

According to the Bruzual \& Charlot (2003) models discussed in $\S$5.3, an
established stellar population will passively fade by $\simeq$1 magnitude in
rest-frame $I$ since $z\simeq$1.2 (e.g. Fontana et al. 2004\nocite{fpd+04}). Accordingly, in
the absence of growth by accretion (Treu et al 2004)\nocite{treu04?}, our survey sensitivity 
corresponds to locating the progenitors of current epoch 0.6L$^{\ast}$ galaxies.
Our inferred abundance of $\phi(L>0.6L^*) = 0.6\phi ^*  =3\times10^{-3}Mpc^{-3}$, 
or just over a half of all galaxies today.

We thus find agreement with Firth et al. (2002) that the ERO population is
sufficient to explain $\sim 50$\% the present-day spheroidal population solely via
passive evolution. This is consistant with purely photometric determinations 
of the evolving red rest-frame luminosity function (e.g. Chen et al. 2004) and
both spectroscopic and photometric measures of the evolving mass density in red
systems (e.g. Glazebrook et al. 2004; Bell et al. 2004; Fontana et al. 2004).
While the most up-to-date semi-analytic hierarchical merging
models (e.g. Sommerville et al. 2004) can produce the required number density of
massive systems at $z \sim 2$, they have great difficulty in reproducing the
high space density of massive galaxies with red colours and evolved stellar populations.

\subsection{Clustering of the ERO Population}
\label{subsec:cluster}

One of the motivations for defining the redshift distribution of the ERO
population is to better understand their spatial clustering. McCarthy et al (2001)
measured the angular clustering of the $I-H>$3 population in the HDF-S and CDF-S
field, a total area of 0.39 deg$^2$. Within the range appropriate
for verification via our study (19$<H<$20.5), they found an amplitude $\theta_0$\,=\,6.7 $\pm$ 0.4 
arcsec for the two-point correlation function (assuming a form $w(\theta)\,=\,(\theta_0\,/\,\theta)^{\gamma-1}$ where the slope of the correlation function $\gamma$ is -1.8 
as in local samples). To invert the angular function and its normalization $\theta_0$
into its spatial equivalent for comparison with local sources, McCarthy et al 
adopted a Gaussian $N(z)$ with a peak at $z$=1.2 and a dispersion $\sigma_z$=0.3 .
They obtained a clustering length $r_0$=9.8 $h^{-1}$ Mpc - a value considered to
be larger than that for present day galaxies but comparable to that for the 
local elliptical population (Norberg et al 2001).

Firth et al (2002) subsequently analyzed the angular clustering more rigorously 
in the HDS-S field. For a $H<$20.5 sample of 170 galaxies with $I-H>3$ across 
a smaller area of 0.21 deg$^2$ a more refined photometric redshift 
distribution was derived. They relate the correlation length $r_0$ at 
a given redshift $z$ to its zero redshift equivalent, say $r_{z0}$, via the 
expression:

$$r_0(z) = r_{z0}\, (1\,+\,z)^{(\gamma-\epsilon-3)/\gamma}$$

where $\epsilon$ parameterizes the evolution of clustering. $\epsilon$=0
corresponds to constant clustering in proper space and $\epsilon$ = -1.2 constant
clustering in comoving coordinates. Assuming their ERO population did not
evolve in comoving terms (i.e. $\epsilon$=-1.2), Firth et al deduced the
population would have a local clustering with $r_{z0}$=7.5 $\pm$3.7, i.e. 1.5
times larger than local galaxies but similar to that observed ($r_{z0}\simeq$6-8
Mpc) for local ellipticals. Most significant of all, the clustering of
the ERO sample in both studies is a factor of 2.5-4 times stronger than the 
full $H$-selected population. The larger error bar on the Firth et al determination 
compared to McCarthy et al is only partly due to the smaller field employed; 
it also reflects a more realistic measure of the overall inversion uncertainty.

The availability of spectroscopic redshifts allows us to strengthen these
conclusions in two ways. Firstly, the improved precision of a spectroscopic
redshift over its photometric equivalent permits us to directly detect
large scale structures composed of EROs rather than relying entirely
on projected angular statistics. Secondly, the deduced redshift distribution,
$N(z)$ (see $\S$4) allows us to significantly improve upon the clustering
analyses undertaken by McCarthy et al and Firth et al.

\begin{figure}
\psfig{figure=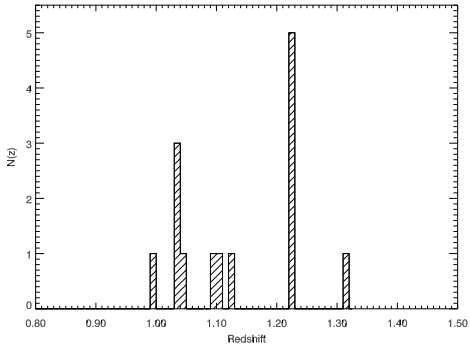,width=50mm,angle=90} \\
\caption{Large Scale Structure in the CDFS (including sources
selected using $R-K>5$). The bin size is $\Delta z=0.02$ (400km~s$^{-1}$).}
\label{fig:cdfs-distrib}
\end{figure}

\begin{figure}
\psfig{figure=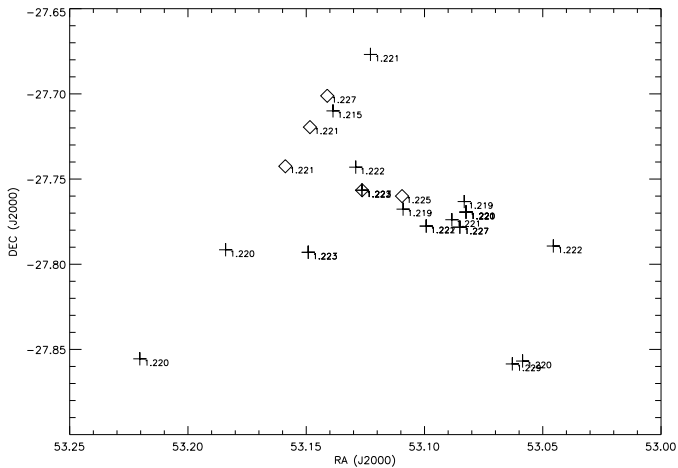,width=65mm,angle=90} \\
\caption{Angular distribution of galaxies around the $z=1.22\pm0.01$
structure in the CDFS. Each galaxy is labelled with its spectroscopic 
redshift. Crosses represent sources in the FORS2 sample, diamonds those 
from the present survey.}
\label{fig:cluster}
\end{figure}

To illustrate the former, our spectroscopy, when combined with those
selected with $R-K>$5 in the FORS2 sample (Vanzella et al (2004) \nocite{van04})
define a prominent overdensity of EROs at $z$=1.22 in the CDFS 
(Figure~\ref{fig:cdfs-distrib}). The angular distribution of 20
sources within $\delta\,z=\pm$0.01 (i.e. a velocity dispersion of $\sim400$~km~s$^{-1}$) is shown in Figure~\ref{fig:cluster} 
and appears to trace a large wall-like structure possibly consistent with 
an assembling cluster.  The five LCIRS galaxies in the structure all 
exhibit signs of a prominent old stellar population, whereas the majority
in the FORS2 sample were spectroscopically identified by [O{\scriptsize~II}] 
emission alone. This discrepancy {\it may} be explained by our selection function for EROs. However,
 assuming that this is a gravitationally bound cluster, we speculate that the
relative spatial concentrations of the two galaxy types
might also be evidence for the morphology-density relation (Dressler et
al.\ 1997 \nocite{doc+97}) operating at higher redshift. In summary,
we are possibly seeing the older, evolved galaxies at the centre of
the cluster and the more active star-forming galaxies tracing the
outskirts. Assuming further that the cluster is virialized, we find a mass
of $\sim5.6\times10^{14}\msol$ within a radius of 5~Mpc. This is consistent with other high redshift clusters , for
example, the rich cluster MS1054-03 which has a mass of $\sim1.9\times10^{15}\msol$ within $\sim2~Mpc$ (Tran et
al. 1999)\nocite{tkd+99}.
                                         
Returning finally to the spatial correlation function, we can
verify more precisely the discussion in Firth et al (2002) by comparing
the photometrically-inferred redshift distribution $N(z)$ for a
$H<20.5$, $I-H>$3.0 sample (Figure 29 in Firth et al's paper) with
our spectroscopic equivalent (Figure 6). As we have already noted,
the distributions are remarkably similar so we can expect little
change in the analysis of Firth et al.

However, taking the angular clustering signal from the larger area 
studied by McCarthy et al (2001), we infer an amplitude for the 
angular correlation function $w(\theta)$ at $\theta$=1.0 arcmin of 
$A$=0.173 $\pm$0.010 (c.f. 0.20 $\pm$ 0.06 from Firth et al's HDF-S
study) and using the redshift distribution of Figure 6, this inverts 
to a local clustering scale length for the comoving case of $r_{z0}$= 
6.5 $\pm$ 0.4. This is somewhat smaller than the amplitude claimed by 
Firth et al but a more precise estimate.

Our measurement is slightly smaller, but potentially more accurate, than most $r_0$ amplitudes in the literature relying
on photometric redshift estimates. For example, Daddi et al. (2002) quote
a range $5.5\aplt r_0/(h^{-1}\, {\rm Mpc})\aplt 16$ (note they use H$_0=100h$) for old passively
evolving EROs. Brown et al. (2005)\nocite{brown2005} estimate the spatial clustering of EROs
over a large area (0.98$^2$\,degrees) in the NOAO Deep Wide Field Survey, using photometric redshifts,
and extrapolate a value of $r_{0}\sim7.5h^{-1}$\,Mpc for $K\apgt 20$, where H$_0=100h$.

\section{CONCLUSIONS}

We have used $H$-band imaging from the LCIRS survey to select a large and 
uniform sample of EROs brighter than $H>20.5$ with a colour selection
$(I-H)>3$, over three fields (SSA22, CDFS \& NTT Deep Field). 

Our deep Keck LRIS \& DEIMOS spectroscopy has targetted 50\% of these EROs
over a sampling area of 200\,arcmin$^{2}$. Of the 67 EROs appearing on
our slitmasks, we have determined reliable spectroscopic redshifts for 44 
-- a completeness fraction of 66\%. This is the most extensive spectroscopic 
study of this population to date, and the first to focus on an $H$-band selection.
Most of our spectra have a continuum signal/noise of 2--3 per \AA\ and 
the superior resolution of our DEIMOS data enables us to undertake diagnostic 
spectroscopic studies of various subsets of the ERO population.

We find that most of the ERO population contain a clear signature of
an old stellar population. In contrast to other studies, based on
smaller samples with weaker signal/noise and lower dispersion data,
only a minority of our sample (16\%) appears to be due to genuinely
young dust-reddened systems. We suggest the discrepancy may arise
in part because of our choice of $I-H$ as a colour discriminant
rather than the coarser $R-K$.

However, within our dominant population of established galaxies, we
notice a great diversity in spectral properties. We classify these 
as pure E (27\%), E+A (13.5\%), E+e (23\%) and 'mixed' E+A+e (13.5\%) depending on the presence
or otherwise of secondary features of star formation such as [O II]
emission and post-burst Balmer absorption. We propose that these 
different subsets represent various manifestations of the same 
overall population seen at different stages as they accrete
associated objects and via illustrative spectral modelling deduce that
the bulk of the stars in the population formed well before $z\simeq$2.

The abundance of the E+A and similar systems in our sample is an extremely
interesting result. The numbers are quite high, in comparison with the
local universe where they are very rare, and imply we may be seeing the
fading light of a recent major star formation epoch.  

As earlier papers in this series have surmised, the abundance and 
luminosity distribution of this established component of EROs,
when faded by passive evolution to the present epoch, cannot account
for the population of local early-type galaxies. As many workers
are deducing (Treu et al 2005, Bundy et al 2005\nocite{bfe+04}), continued 
growth and transformations are required over 0.5$<z<$1.5 to match
the local distribution. However, as many authors have claimed, the 
evolutionary growth in the number of early-types since $z<4$ is far 
less dramatic than that predicted in recent semi-analytical models.

We finally refine earlier estimates of the spatial clustering of
this population taking into account the improved redshift distribution
in the inversion from the angular clustering seen in the overall
LCIRS survey. We find a spatial correlation length, corrected to
the present epoch assuming no evolution in comoving coordinates,
of $r_0$=6.5 $\pm$ 0.4 $h^{-1}$ Mpc, a signal comparable to but
marginally less than that seen in present day ellipticals.

\subsection*{ACKNOWLEDGMENTS}

We acknowledge the LCIRS collaboration for the provision of an unique
database upon which this detailed spectroscopic study was based.
In particular, we are indebted to Hsiao-Wen Chen, Andrew Firth, Richard
McMahon, Chris Sabbey and Ofer Lahav for their important contributions.
The CIRSI camera was made possible by the generous support of
the Raymond and Beverly Sackler Foundation. 

We also thank various individuals in the Keck community for their
assistance. We thank Sandy Faber, Judy Cohen, Chuck Steidel and the 
DEIMOS/LRIS teams for making these impressive instruments a reality.
The software used to design the slitmasks was written by Drew
Phillips. The LRIS slitmask data reduction IRAF package, BOGUS, was 
written by Daniel  Stern, S.\ Adam Stanford \& Andrew Bunker. The 
DEIMOS data analysis pipeline was developed at UC Berkeley with support 
from NSF grant AST-0071048. We are grateful to Michael Cooper \& Alison Coil 
for their advice in the use of this package.

Data presented herein were obtained at the W~ M.\ Keck
Observatory, which is operated as a scientific partnership among the
California Institute of Technology, the University of California, and
the National Aeronautics and Space Administration. The Observatory was
made possible by the generous financial support of the W.~M.\ Keck
Foundation. 

We thank Stephane Charlot \& Gustavo Bruzual for making
their useful stellar population synthesis code available. MD is 
grateful for support from the Fellowship Fund Branch of AFUW Qld Inc., 
the Isaac Newton Studentship, the Cambridge Commonwealth Trust and 
the University of Sydney.

Finally, we thank the referee, Andrea Cimatti, for detailed constructive comments which
have improved the paper, and Karl Glazebrook, for some helpful comments on the manuscript.

\bibliographystyle{/home/md/LaTeX/mn2e}
\bibliography{/home/md/LaTeX/BibTeX/myrefs}

\newpage

\begin{figure*}
\psfig{figure=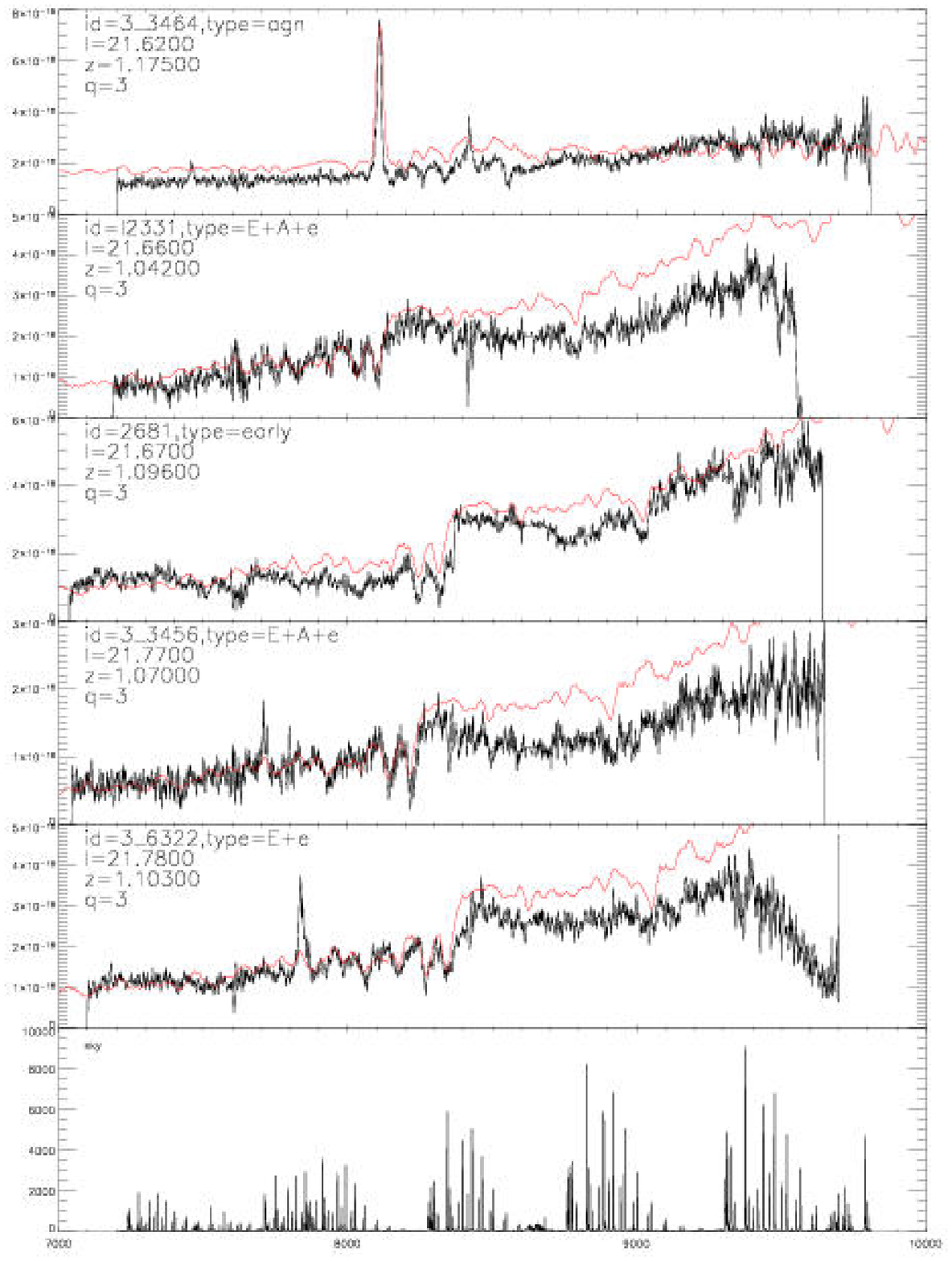,height=260mm} \\
\caption{Individual DEIMOS spectra ordered by $I$ magnitude from bright 
to faint and smoothed to an effective resolution of 3.5 \AA\ .}
\label{fig:deimos_spec}
\end{figure*}

\begin{figure*}
\psfig{figure=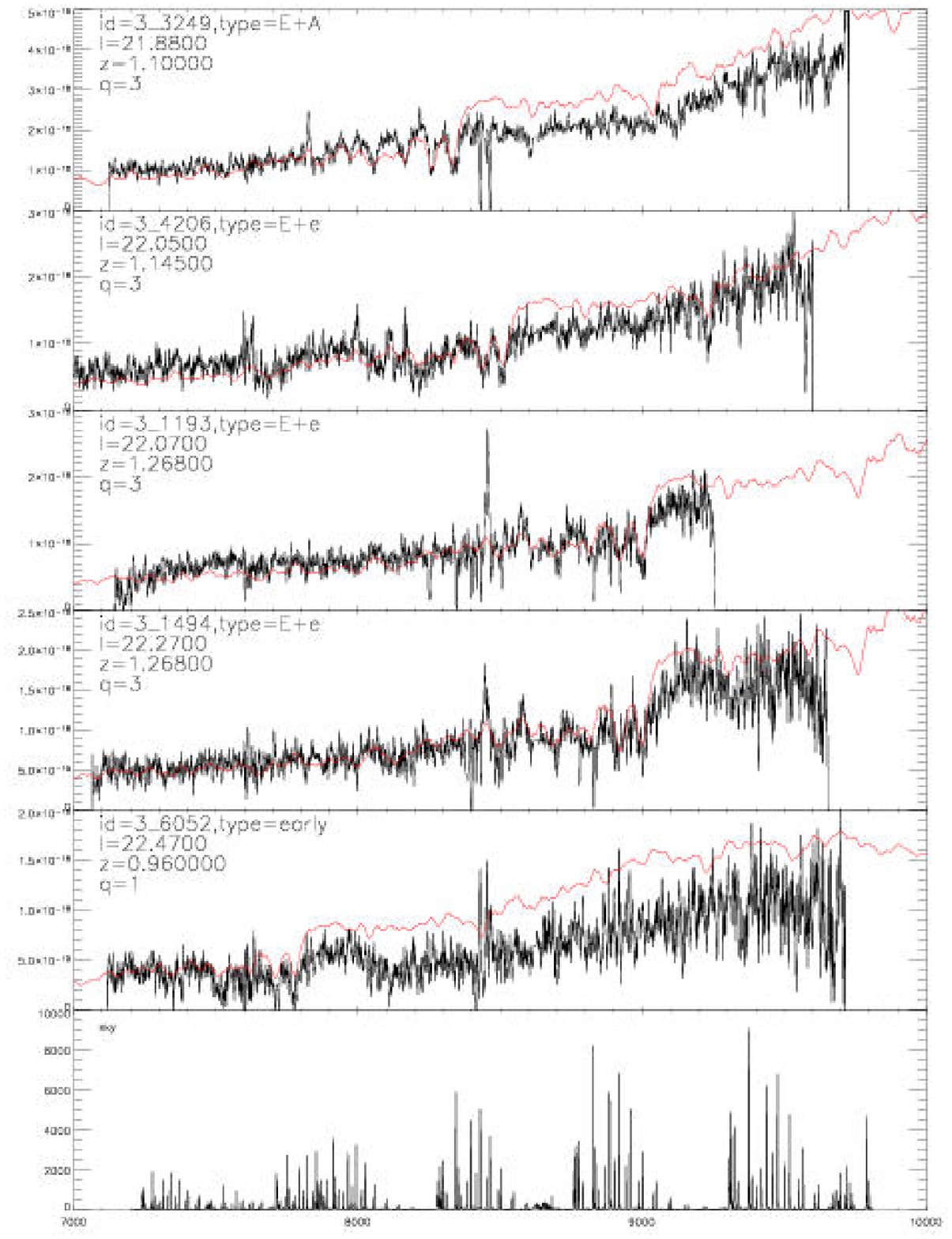,height=260mm}\\ 
\end{figure*}
\begin{figure*}
\psfig{figure=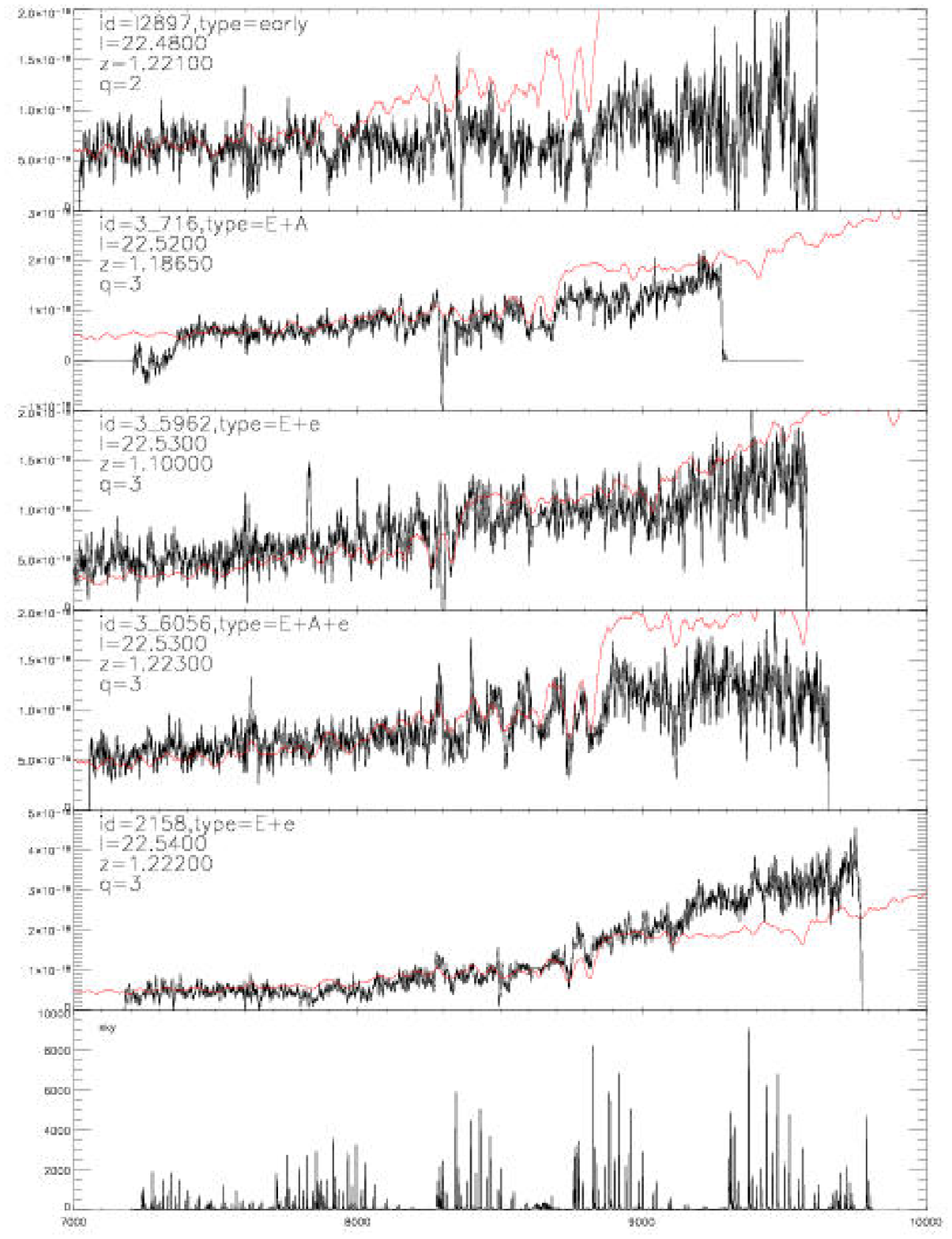,height=260mm}\\  
\end{figure*}
\begin{figure*}
\psfig{figure=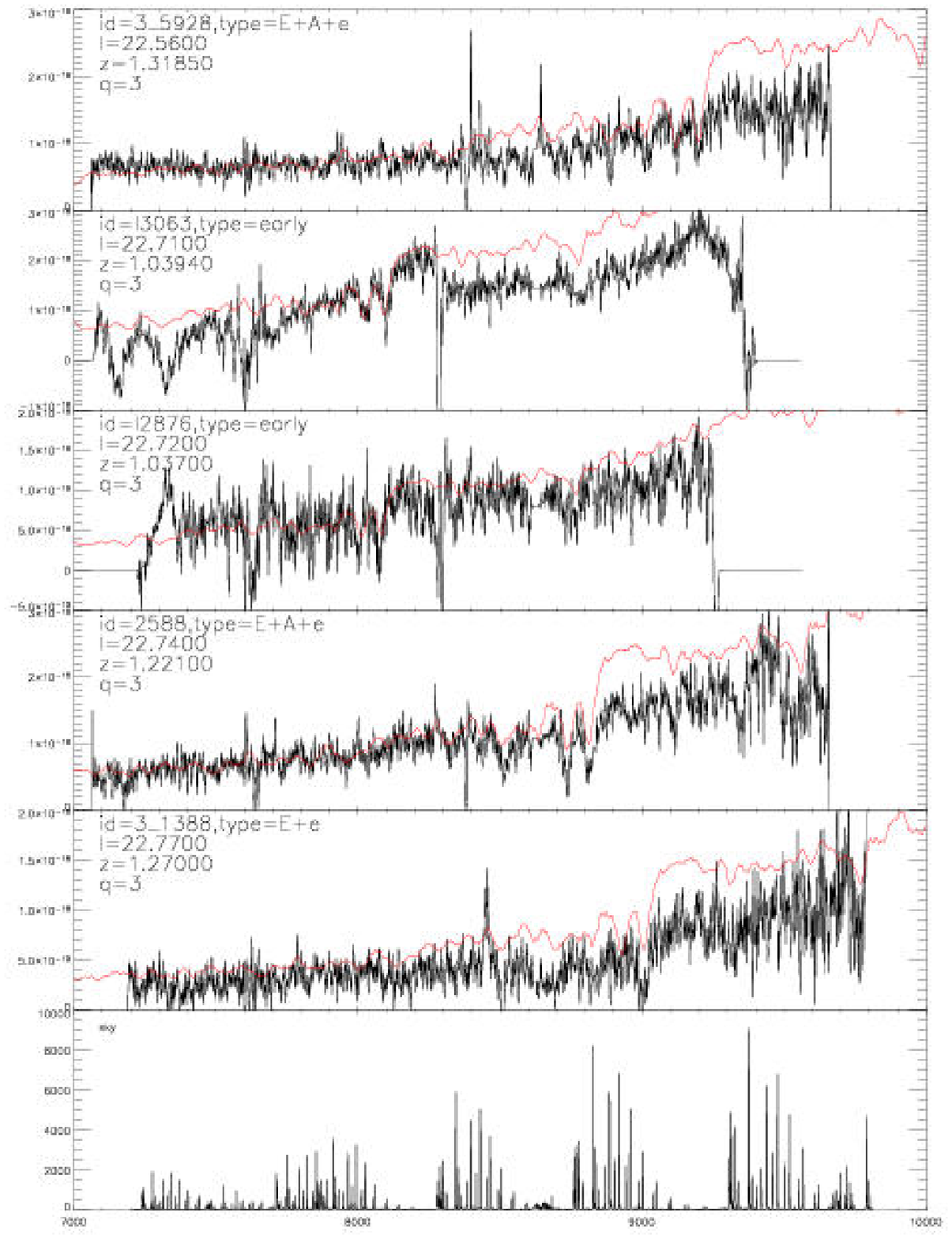,height=260mm}\\ 
\end{figure*}
\begin{figure*}
\psfig{figure=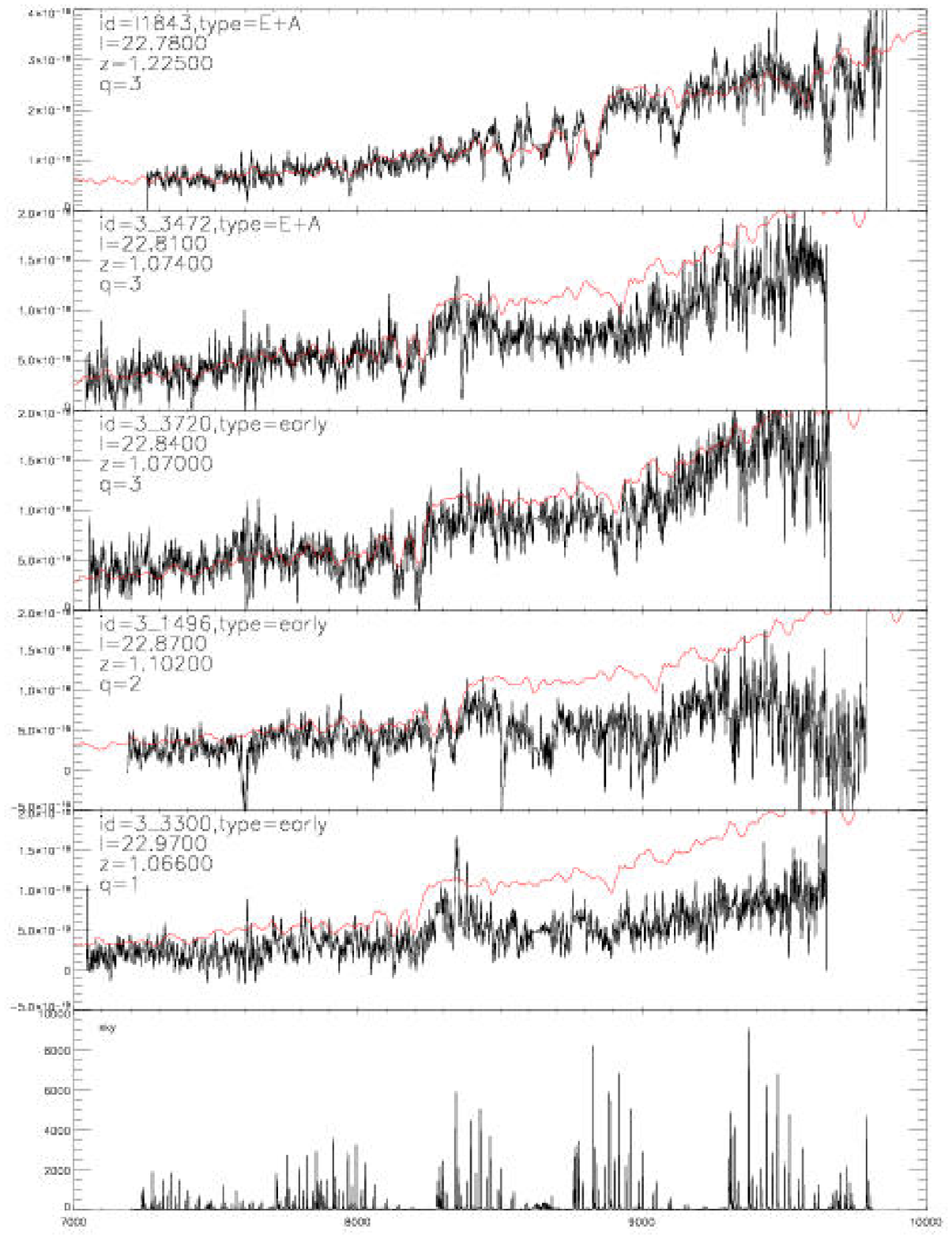,height=260mm}\\ 
\end{figure*}
\begin{figure*}
\psfig{figure=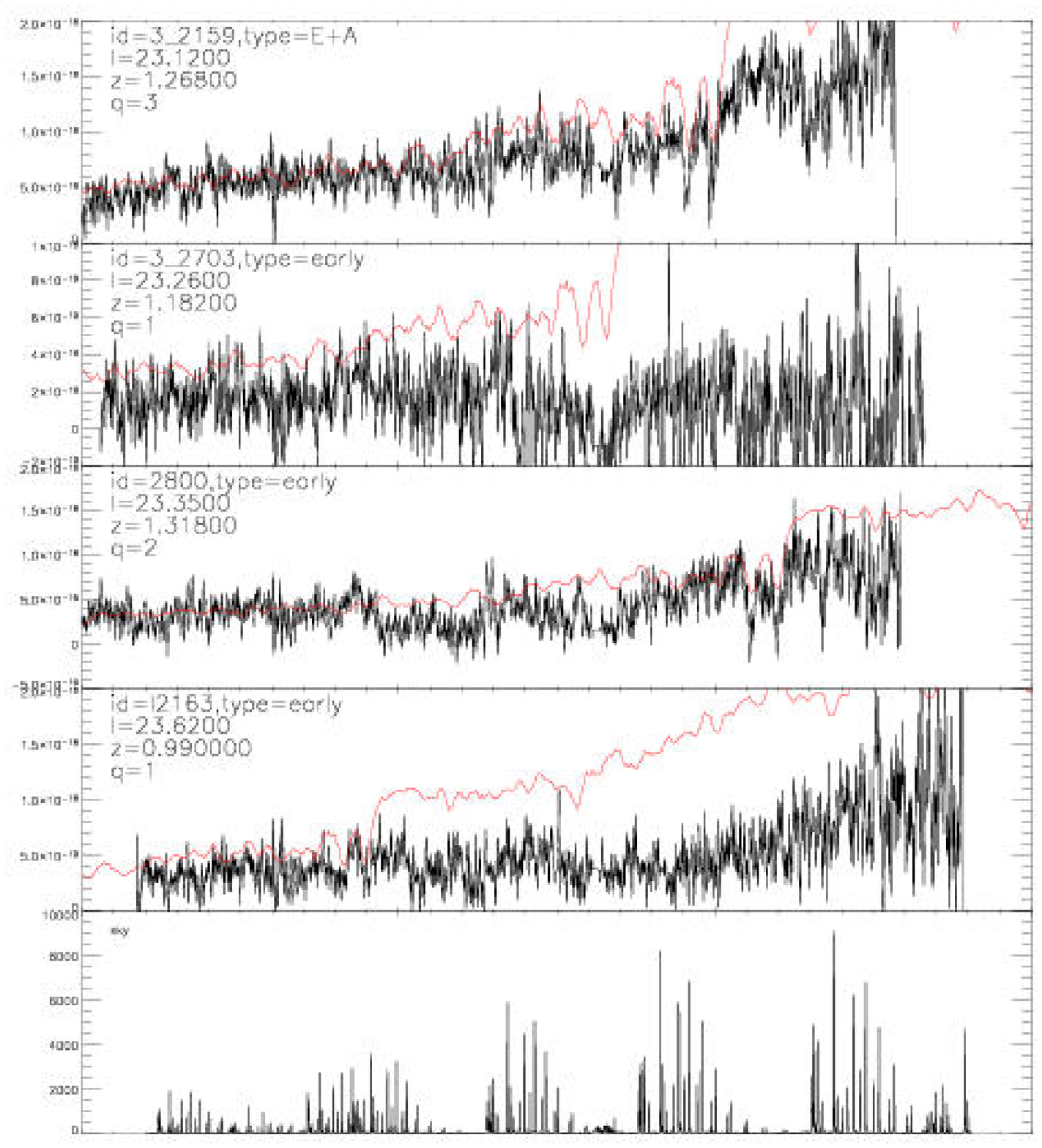,height=260mm}\\ 
\end{figure*}

\begin{figure*}
\psfig{figure=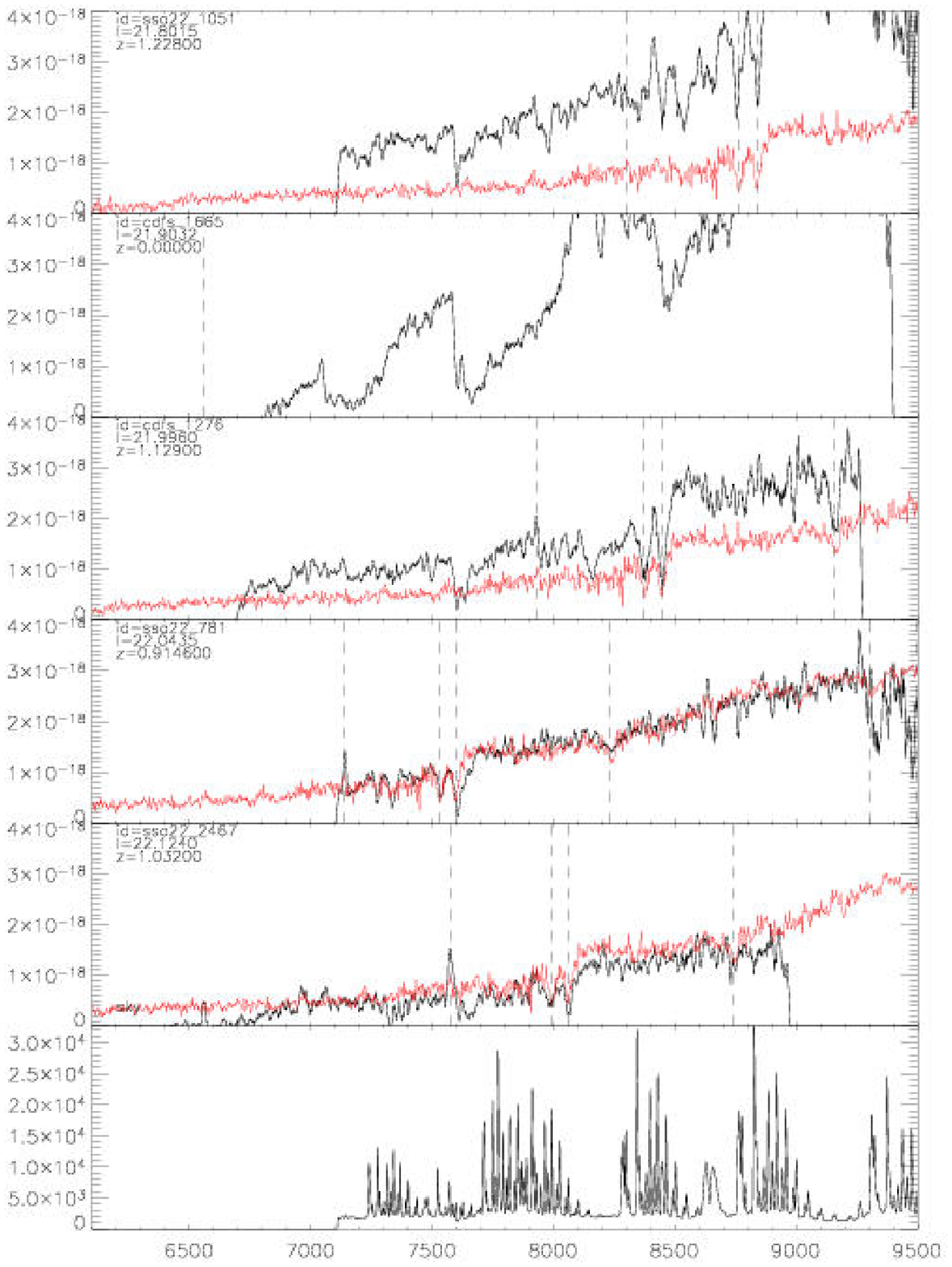,height=260mm} \\
\caption{Individual LRIS spectra ordered by $I$ magnitude from bright to
faint and smoothed to an effective resolution of 13.6 \AA\ }
\label{fig:deimos_spec}
\end{figure*}%

\begin{figure*}
\psfig{figure=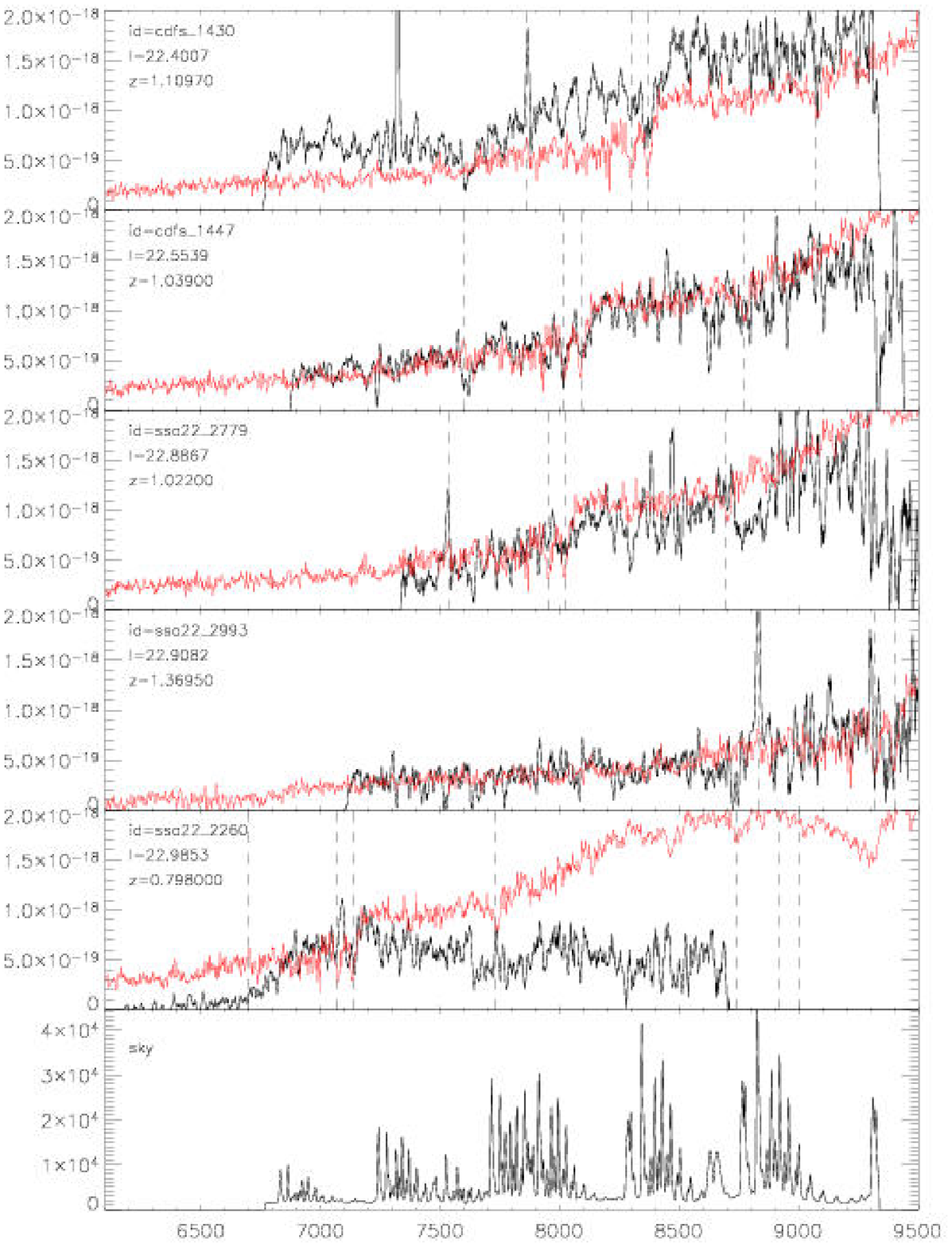,height=260mm}\\ 
\end{figure*}
\begin{figure*}
\psfig{figure=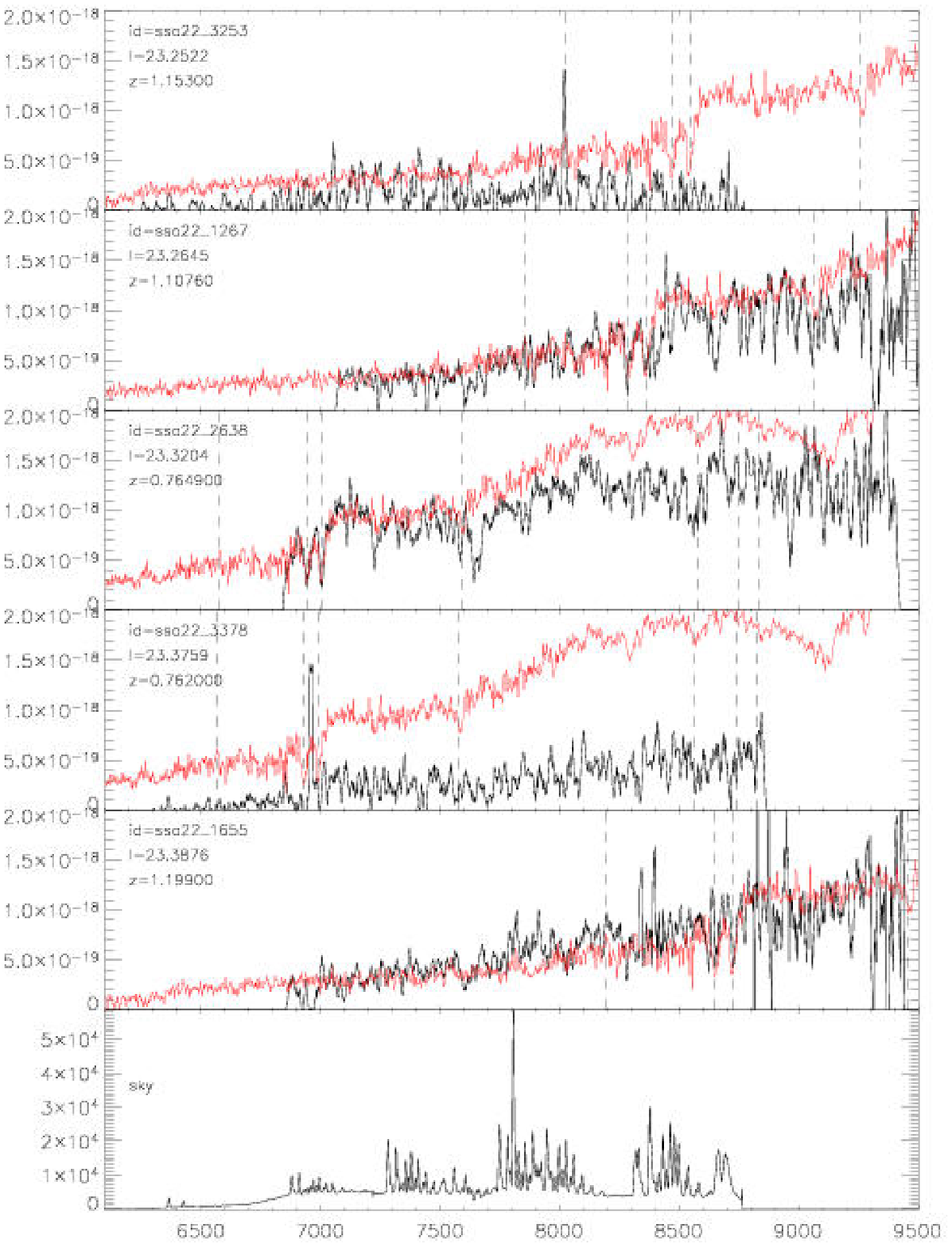,height=260mm}\\  
\end{figure*}

\label{lastpage}

\end{document}